\documentclass[article]{JFM-FLM_Au_mod}
\usepackage{mathtools}
\usepackage{svg}
\usepackage{subcaption}
\usepackage[dvipsnames]{xcolor}
\usepackage{graphicx}
\usepackage{hyperref}
\hypersetup{colorlinks=true,citecolor=blue}

\lefttitle{L. Meissner, B. Cichocki and J.C.Everts}
\righttitle{Journal of Fluid Mechanics}

\title{Exact results for dissipation and steady creeping flow in three-dimensional chiral active fluids}

\author{Laura Meissner-Oszer\aff{1}, Bogdan Cichocki\aff{1} \and Jeffrey C. Everts\aff{1,2}}

\affiliation{\aff{1}Institute of Theoretical Physics, Faculty of Physics, University of Warsaw, Pasteura 5, 02-093 Warsaw, Poland
\aff{2}Institute of Physical Chemistry, Polish Academy of Sciences, 01-224 Warsaw, Poland}

\corresau{Jeffrey C. Everts, \email{jeffrey.everts@fuw.edu.pl}}

\def\*#1{\mathbf{#1}}
\def\!#1{\mathbf{\hat#1}}

\begin{document}

\maketitle
\thispagestyle{empty}

\begin{abstract}
Chiral active fluids consist of self-spinning particles that rotate due to continuous energy injection at the microscopic scale (e.g., by activity or an external field). The hydrodynamics of such fluids is described by antisymmetric contributions in the viscosity tensor --called odd viscosity--, which are allowed by symmetry due to the presence of a non-trivial spin angular momentum density. By generalising the Helmholtz minimum dissipation theorem to systems with odd viscosity, we show that incompressible three-dimensional odd fluids in the presence of sources that induce flow (e.g. surfaces that impose boundary conditions) admit a unique solution for their steady flow fields at low Reynolds number. Furthermore, we prove that such flows dissipate more energy than ordinary Stokes flow, provided that the flow field is affected by odd viscosity. As an example, we consider a model fluid described by one shear viscosity and one odd viscosity in the creeping flow regime. We explicitly compute the stress tensor for a fluid subjected to a point force density. Finally, we compute exact results for the pressure and flow fields around a translating and rotating spherical particle from their singularity representations. From these solutions and our extended Helmholtz theorem, we explain why a translating sphere dissipates more energy when odd viscosity is present, whereas a rotating sphere does not.
\end{abstract}



\section{Introduction}
\label{sec:introduction}
Intrinsically rotating structures are recurring features in driven systems such as molecular motors \citep{Sumino:2012}, cilia in living organisms \citep{Cartwright:2004}, sperm cell colonies \citep{Riedel:2005}, and artificial robotic particles \citep{Scholz:2021}-- all of which display fluid-like behaviour over mesoscopic to macroscopic scales. The effective hydrodynamics of these so-called chiral active fluids is governed not only by the transport of mass and linear momentum, but also by the dynamics of spin angular momentum. The presence of this additional degree of freedom fundamentally modifies the constitutive relations between stresses and strain rates, allowing for antisymmetric components in the viscosity tensor, commonly referred to as odd viscosity \citep{Avron:1988, Fruchart:2023}. Such terms are permitted by the Onsager–Casimir reciprocity relations \citep{Onsager:1931a, Onsager:1931b, Casimir:1945} when the orientation of the intrinsic spin angular momentum is included as an additional degree of freedom. Odd viscosity can lead to unconventional flow patterns \citep{Khain:2022, Lier:2024}, affect turbulence \citep{Chen:2024,Wit:2024}, and, in certain cases, endow the fluid with topological properties \citep{Souslov:2019, Lou:2022}. Although odd viscosity has been recognized for decades --albeit known under different names such as transverse viscosity \citep{Beenakker:1971} and gyroviscosity \citep{Callen:1992}-- recent discoveries have sparked a renewed interest in the study of odd fluids. Notable examples of such studies are the experimental realisation of odd viscosity in graphene \citep{Geim:2019} and in a colloidal fluid of magnetic cubes \citep{Soni:2019}.

Biology and soft matter provide various possibilities for chiral active fluids. Due to the length scales involved in such systems, the creeping-flow (low Reynolds number) regime is of special interest, and has recently been the subject of intense study. In particular, (quasi) two-dimensional odd systems have been investigated for their flow properties and resulting drag forces \citep{Ganeshan:2017, Lier:2023, Hosaka:2025}. In three-dimensional odd fluids, work has, for example, focused on a full classification of the types of odd viscosities \citep{Khain:2022}, odd Stokesian dynamics \citep{Cruz:2023}, microswimmers suspended in odd fluids \citep{Hosaka:2024}, the Lorentz reciprocal theorem \citep{Vilfan:2023}, and the motion of suspended microparticles of various shapes \citep{Khain:2024}. Furthermore, the full Green's function of an unbounded odd fluid has been computed analytically, including the full grand mobility matrix of a sphere suspended in it \citep{Everts:2024,Everts:2024b}. 

Despite significant recent progress, our understanding of microhydrodynamics in three-dimensional chiral active fluids remains limited compared with the extensive results available for ordinary Stokesian fluids \citep{HappelBrenner,Kim}. Here, we address this gap by deriving exact results for steady incompressible three-dimensional chiral active flows at low Reynolds number. First, we establish a theorem for viscous dissipation in odd fluids by generalizing the Helmholtz theorem \citep{Helmholtz}, which expands upon our previous analysis of the viscous dissipation due to a translating and rotating passive sphere in a fluid with odd viscosity \citep{Everts:2024,Everts:2024b}. Second, we compute the stress response to a localised point-force density, a quantity frequently required in computational techniques such as the boundary element method \citep{pozrikidis1992}. Third, we provide an intuitive derivation for the singularity representation of the flow and pressure fields around suspended (solid) particles, complementing more formal and rigorous analyses \citep{Brenner:1964b,Brenner:1964,Brenner:1966}. Finally, we derive exact closed-form expressions for the flow and pressure around a translating and rotating sphere in a chiral active fluid using the singularity representations. For the translating sphere, we also perform an in-depth analysis of the flow streamlines for arbitrary directions of the fluid's spin momentum. Here, we go beyond the fluid-flow results from \citet{Everts:2024,Everts:2024b} where  closed-form expressions were only presented for spheres translating parallel to the fluid's spin-angular momentum.  Finally, we  show that our results are fully consistent with our generalized Helmholtz theorem.

\section{Viscous dissipation in systems with odd viscosity}
For steady Newtonian fluids in the creeping-flow regime, two fundamental results emerge from analysing the viscous dissipation rate. The first result concerns the uniqueness of the solution to the incompressible Stokes equations. The second states that among all incompressible (i.e., solenoidal) flows under specified boundary conditions, the flow satisfying the linear momentum balance has the minimal dissipation rate \citep{Kim}. The latter result is known as the Helmholtz minimum dissipation theorem. We aim to generalize these key results to arbitrary stress–strain constitutive relations, including those for chiral active fluids, to assess the role of odd viscosity in viscous dissipation.

Consider an incompressible chiral active fluid with a spatially constant spin angular-momentum density $\boldsymbol{\ell}$, fluid velocity $\boldsymbol{v}(\boldsymbol{r})$, and pressure $p(\boldsymbol{r})$. The corresponding stress tensor is $\boldsymbol{\sigma}(\boldsymbol{r})=-p(\boldsymbol{r})\mathsfbi{I}+\boldsymbol{\sigma}^\mathrm{V}(\boldsymbol{r})$. In the framework of linear irreversible thermodynamics, the viscous part of the stress tensor satisfies the constitutive relation ${\boldsymbol{\sigma}^\mathrm{V}(\boldsymbol{r})=\boldsymbol{\eta}:\nabla\boldsymbol{v}(\boldsymbol{r})}$, with $\boldsymbol{\eta}$ being the viscosity tensor. Here, we define the double contraction as $[\mathsfbi{A}:\mathsfbi{B}]_{\alpha...\beta} = A_{\alpha...\lambda\nu}B_{\nu\lambda...\beta}$, with summation implied over repeated indices. Greek indices range over all Cartesian coordinates and, for the moment, we do not put any restriction on the spatial dimension. In this work, we also make use of the single contraction defined as $[\mathsfbi{A}\cdot\mathsfbi{B}]_{\alpha...\beta} = A_{\alpha...\lambda}B_{\lambda...\beta}$ and we use carets to denote unit vectors, e.g. for an arbitrary vector $\boldsymbol{w}$, we write $\hat{\boldsymbol{w}}=\boldsymbol{w}/|\boldsymbol{w}|$. Microscopic time reversibility constrains the form of $\boldsymbol{\eta}$ via the Onsager-Casimir reciprocal relations $\eta_{\alpha\beta\lambda\nu}({\boldsymbol{\ell}})=\eta_{\lambda\nu\alpha\beta}(-{\boldsymbol{\ell}})$ \citep{Groot:1964, Casimir:1945}. 

The microscopic origin for a non-zero $\boldsymbol{\ell}$ in chiral active fluids requires the presence of active sources of torque in the fluid. However, \citet{Markovich:2021} have shown that when $\boldsymbol{\sigma}^\mathrm{V}$ is interpreted as the viscous stress tensor for the total linear momentum, that it necessarily has to be symmetric, i.e. $\sigma_{\alpha\beta}^\mathrm{V}(\boldsymbol{r})=\sigma_{\beta\alpha}^\mathrm{V}(\boldsymbol{r})$. An alternative justification is provided by \citet{Banerjee:2017} who interpret $\boldsymbol{\sigma}^\mathrm{V}$ as the stress tensor for the centre-of-mass momentum. They show that antisymmetric stresses can be neglected when $\boldsymbol{\ell}$ acquires a spatially constant value in a non-equilibrium steady state. Therefore, we assume that there are \emph{effectively} no intrinsic sources of torque in the fluid. It follows that ${\boldsymbol{\sigma}^\mathrm{V}(\boldsymbol{r})=\boldsymbol{\eta}(\boldsymbol{\ell}):\mathsfbi{e}(\boldsymbol{r})}$ with rate of strain tensor $e_{\alpha\beta}(\boldsymbol{r})=[\partial_\alpha v_\beta(\boldsymbol{r})+\partial_\beta v_\alpha(\boldsymbol{r})]/2$. Furthermore, from the incompressibility condition it follows that $\eta_{\lambda\lambda\alpha\beta}(\boldsymbol{\ell})=0$.  Combining this with the Onsager-Casimir relations, we obtain  $\eta_{\alpha\beta{\lambda}\lambda}(\boldsymbol{\ell})=0$. 

Let $\mathcal{V}$ be a fluid domain with boundary $\mathcal{S}$ with stick boundary condition on all surfaces. Taking the above symmetry considerations into account, the equations governing steady fluid flow at low Reynolds number are given by the balance of linear momentum and the incompressibility condition
\begin{equation}
\nabla\cdot[\boldsymbol{\eta}(\boldsymbol{\ell}):\mathsfbi{e}(\boldsymbol{r})]-\nabla p(\boldsymbol{r})=-\boldsymbol{f}(\boldsymbol{r}), \quad \nabla\cdot\boldsymbol{v}(\boldsymbol{r})=0, \quad \boldsymbol{r}\in\mathcal{V}, \label{eq:creeping}
\end{equation}
respectively, with $\boldsymbol{v}(\boldsymbol{r})$ given for $\boldsymbol{r}\in\mathcal{S}$ and $\boldsymbol{f}(\boldsymbol{r})$ a body-force density. We decompose the viscosity tensor for a general odd fluid as $\boldsymbol{\eta}=\boldsymbol{\eta}^\mathrm{S}+\boldsymbol{\eta}^\mathrm{A}$, where $\eta_{\alpha\beta\lambda\nu}^\mathrm{S}(\boldsymbol{\ell})=\eta_{\lambda\nu\alpha\beta}^\mathrm{S}(\boldsymbol{\ell})$ and $\eta_{\alpha\beta\lambda\nu}^\mathrm{A}(\boldsymbol{\ell})=-\eta_{\lambda\nu\alpha\beta}^\mathrm{A}(\boldsymbol{\ell})$.  The tensor $\boldsymbol{\eta}^\mathrm{A}$ contains all the odd-viscosity coefficients.
The total dissipated power due to viscous effects is  
\begin{equation}
\dot{E}=\int_\mathcal{V}\mathrm{d}V\, \boldsymbol{\sigma}^\mathrm{V}(\boldsymbol{r}):\nabla\boldsymbol{v}(\boldsymbol{r})=\int_\mathcal{V}\mathrm{d}V\,\mathsfbi{e}(\boldsymbol{r}):\boldsymbol{\eta}^\mathrm{S}:\mathsfbi{e}(\boldsymbol{r}). \label{eq:disspower}
\end{equation}
Note that for systems in local thermodynamic equilibrium (implied in this construction), $\mathsfbi{e}(\boldsymbol{r}):\boldsymbol{\eta}^\mathrm{S}:\mathsfbi{e}(\boldsymbol{r})\geq 0$ holds for any $\mathsfbi{e}(\boldsymbol{r})$, which follows from the second law of thermodynamics \citep{Groot}.

\subsection{Uniqueness of solutions}
\label{sec:unique}
Let $(\boldsymbol{v},p)$ and $(\boldsymbol{v}',p')$ be two solutions of Eq. \eqref{eq:creeping} with the same boundary condition, $\boldsymbol{v}'(\boldsymbol{r})=\boldsymbol{v}(\boldsymbol{r})$ for $\boldsymbol{r}\in\mathcal{S}$. Furthermore, denote their corresponding stress tensors and strain rate tensors as $(\boldsymbol{\sigma},\mathsfbi{e})$ and $(\boldsymbol{\sigma}',\mathsfbi{e}')$, respectively. Consider the following steps for the viscous dissipation of their difference fields:
\begin{align}
   \Delta \dot{E}&:=
   \int_\mathcal{V}\mathrm{d}V\,[e_{\alpha\beta}'(\boldsymbol{r})-e_{\alpha\beta}(\boldsymbol{r})]\eta_{\alpha\beta\lambda\nu}^\mathrm{S} [e_{\lambda\nu}'(\boldsymbol{r})-e_{\lambda\nu}(\boldsymbol{r})] \nonumber\\
   &=   \int_\mathcal{V}\mathrm{d}V\,[\partial_\alpha v'_{\beta}(\boldsymbol{r})-\partial_\alpha v_{\beta}(\boldsymbol{r})]\underbrace{\{\eta_{\alpha\beta\lambda\nu} [e_{\lambda\nu}'(\boldsymbol{r})-e_{\lambda\nu}(\boldsymbol{r})]-[p'(\boldsymbol{r})-p(\boldsymbol{r})]\delta_{\alpha\beta}\}}_{=\sigma'_{\alpha\beta}(\boldsymbol{r})-\sigma_{\alpha\beta}(\boldsymbol{r})}\nonumber\\
   &=\int_\mathcal{S}\mathrm{d}S\, \hat{n}_\alpha[ v'_{\beta}(\boldsymbol{r})- v_{\beta}(\boldsymbol{r})][\sigma_{\alpha\beta}'(\boldsymbol{r})-\sigma_{\alpha\beta}(\boldsymbol{r})]\nonumber\\
   &\quad-\int_\mathcal{V}\mathrm{d}V\,[v'_{\beta}(\boldsymbol{r})- v_{\beta}(\boldsymbol{r})][\partial_\alpha\sigma_{\alpha\beta}'(\boldsymbol{r})-\partial_\alpha\sigma_{\alpha\beta}(\boldsymbol{r})]=0, \label{eq:unique}
\end{align}
with $\boldsymbol{\hat{n}}$ a unit normal pointing out of the fluid.
The second equality follows from the incompressibility and the symmetric-stress conditions, and $(\mathsfbi{e}'-\mathsfbi{e}):\boldsymbol{\eta}^\mathrm{A}:(\mathsfbi{e}'-\mathsfbi{e})=0$. The third equality follows from a partial integration and the last equality follows from the equality of $\boldsymbol{v}$ and $\boldsymbol{v}'$ on $\mathcal{S}$ and using Eq. \eqref{eq:creeping}. We conclude that $\mathsfbi{e}'=\mathsfbi{e}$ and from the boundary condition it follows that $\boldsymbol{v}'=\boldsymbol{v}$ throughout $\mathcal{V}$. Therefore, we have shown that Eq. \eqref{eq:creeping} admits a unique solution for $\boldsymbol{v}$. Our uniqueness theorem generalizes the result of \citet{Lapa:2014}, which was restricted to two-dimensional flows with a particular form of $\boldsymbol{\eta}$.

\subsection{Helmholtz minimum dissipation theorem for anisotropic odd fluids}
\label{sec:diss}

From Eq. \eqref{eq:disspower}, it is often mistakenly concluded that odd viscosity does not contribute to viscous dissipation. This is, however, incorrect, because generally speaking $\mathsfbi{e}(\boldsymbol{r})$ can depend on the components of $\boldsymbol{\eta}^\mathrm{A}$. In fact, it was shown for a passive translating sphere in an odd fluid, that odd viscosity leads to higher viscous dissipation than in ordinary Stokes flow \citep{Everts:2024}. Here, we will make statements for the general case. 

The standard Helmholtz minimum dissipation theorem compares the viscous dissipation rate of an incompressible flow that satisfies the linear momentum balance with that of an arbitrary divergence-free flow field under the same boundary conditions. An analogous approach can be used to compare incompressible flows with and without odd viscosity, provided that their even viscosity components are identical. Consider a solution $(\boldsymbol{v},p)$ to Eq. \eqref{eq:creeping} and let $(\boldsymbol{v}^{(0)},p^{(0)})$ be a reference system without odd viscosity which solves Eq. \eqref{eq:creeping} with $\boldsymbol{\eta}^\mathrm{A}=0$ and corresponding viscous dissipation rate $\dot{E}^{(0)}$. Furthermore, $\boldsymbol{v}^{(0)}=\boldsymbol{v}$ on $\mathcal{S}$. Denote the strain rate tensor and stress tensor by $(\boldsymbol{\sigma},\mathsfbi{e})$ and $(\boldsymbol{\sigma}^{(0)},\mathsfbi{e}^{(0)})$, respectively. Following similar steps as the ones leading to Eq. \eqref{eq:unique}, we find
\begin{gather}
\int_\mathcal{V}\mathrm{d}V\, \left[e_{\alpha\beta}(\boldsymbol{r})-e_{\alpha\beta}^{(0)}(\boldsymbol{r})\right]\eta_{\alpha\beta\lambda\nu}^\mathrm{S}e_{\lambda\nu}^{(0)}(\boldsymbol{r})=0. \label{eq:lemma}
\end{gather}
Next, we compute the viscous energy dissipation 
\begin{align}\allowdisplaybreaks
\dot{E}&=\int_\mathcal{V}\mathrm{d}V\, e_{\alpha\beta}(\boldsymbol{r})\eta_{\alpha\beta\lambda\nu}^\mathrm{S}e_{\lambda\nu}(\boldsymbol{r})\nonumber\\
&=\int_\mathcal{V}\mathrm{d}V\,\eta_{\alpha\beta\lambda\nu}^\mathrm{S}\left[e_{\alpha\beta}(\boldsymbol{r})e_{\lambda\nu}(\boldsymbol{r})+e_{\alpha\beta}^{(0)}(\boldsymbol{r})e_{\lambda\nu}^{(0)}(\boldsymbol{r})-e_{\alpha\beta}^{(0)}(\boldsymbol{r})e_{\lambda\nu}^{(0)}(\boldsymbol{r})\right] \nonumber\\
&\stackrel{\eqref{eq:lemma}}{=}\int_\mathcal{V}\mathrm{d}V\,\eta_{\alpha\beta\lambda\nu}^\mathrm{S}\left\{e_{\alpha\beta}^{(0)}(\boldsymbol{r})e_{\lambda\nu}^{(0)}(\boldsymbol{r})+\left[e_{\alpha\beta}(\boldsymbol{r})-e_{\alpha\beta}^{(0)}(\boldsymbol{r})\right]\left[e_{\lambda\nu}(\boldsymbol{r})-e_{\lambda\nu}^{(0)}(\boldsymbol{r})\right]\right\} \nonumber\\
&=\dot{E}^{(0)}+\underbrace{\int_\mathcal{V}\mathrm{d}V\,
\left[e_{\alpha\beta}(\boldsymbol{r})-e_{\alpha\beta}^{(0)}(\boldsymbol{r})\right]\eta_{\alpha\beta\lambda\nu}^\mathrm{S}\left[e_{\lambda\nu}(\boldsymbol{r})-e_{\lambda\nu}^{(0)}(\boldsymbol{r})\right]}_{\geq 0}\geq \dot{E}^{(0)}. \label{eq:Bogdan}
\end{align}
Equality is achieved only when $\mathsfbi{e}=\mathsfbi{e}^{(0)}$. We conclude that adding odd viscous effects to a system with given boundary conditions will always increase viscous dissipation, unless the velocity fields are unaffected by odd viscosity. However, in the latter case, pressure fields can differ: see the example discussed in Sec. \ref{sec:rotsphere}.

\section{Description of a model fluid with odd viscosity and stress response}
\label{sec:hyd}
 The simplest $\boldsymbol{\eta}$ with non-trivial $\boldsymbol{\ell}$ that satisfies the Onsager-Casimir symmetry and which produces a symmetric stress tensor is
\begin{eqnarray}
\eta_{\alpha\beta\lambda\nu}(\boldsymbol{\hat{\ell}})&=&\eta_\mathrm{s}\left(\delta_{\alpha\lambda}\delta_{\nu\beta}+\delta_{\alpha\nu}\delta_{\lambda\beta}-\frac{2}{3}\delta_{\alpha\beta}\delta_{\lambda\nu}\right)\nonumber\\
&&-\eta_\mathrm{o}\hat{\ell}_\kappa(\epsilon_{\kappa\alpha\lambda}\delta_{\nu\beta}+\epsilon_{\kappa\alpha\nu}\delta_{\lambda\beta}+\epsilon_{\kappa\beta\lambda}\delta_{\nu\alpha}+\epsilon_{\kappa\beta\nu}\delta_{\lambda\alpha}). \label{eq:visctensor}
\end{eqnarray}
Here, $\eta_\mathrm{s}$ is the dynamic shear viscosity, and $\eta_\mathrm{o}$ is the odd viscosity-coefficient that quantifies the magnitude of $\boldsymbol{\mathrm{\ell}}$ through the relation $\boldsymbol{\ell} = 4\eta_\mathrm{o}\boldsymbol{\hat\ell}$, following the convention of \citet{Markovich:2021}. The tensors $\delta_{\alpha\beta},\epsilon_{\alpha\beta\nu}$ represent the Kronecker delta and the Levi-Civita symbol, respectively. The corresponding stress tensor is
\begin{equation}
\sigma_{\alpha\beta}(\boldsymbol{r})=-p(\boldsymbol{r})\delta_{\alpha\beta}+2\eta_\mathrm{s}e_{\alpha\beta}(\boldsymbol{r})-2\eta_\mathrm{o}\hat{\ell}_\kappa[\epsilon_{\kappa\alpha\lambda}e_{\lambda\beta}(\boldsymbol{r})+\epsilon_{\kappa\beta\lambda}e_{\lambda\alpha}(\boldsymbol{r})],
\end{equation}
which depends on the fluid velocity solely through $\mathsfbi{e}(\boldsymbol{r})$. Inserting Eq. \eqref{eq:visctensor} in Eq. \eqref{eq:creeping}, we find
\begin{equation}
\eta_\mathrm{s}\nabla^2\boldsymbol{ v}(\boldsymbol{r})-\nabla\tilde p(\boldsymbol{r})+\eta_\mathrm{o}(\boldsymbol{\hat{\ell}}\cdot\nabla)[\nabla\times\boldsymbol{ v}(\boldsymbol{r})]=-\boldsymbol{ f}(\boldsymbol{r}), \quad \nabla\cdot\boldsymbol{ v}(\boldsymbol{r})=0. \label{eq:hyd}
\end{equation}
Here, we introduced the effective pressure $\tilde{p}(\boldsymbol{r})=p(\boldsymbol{r})+2\eta_\mathrm{o}\boldsymbol{\hat{\ell}}\cdot[\nabla\times\boldsymbol{ v}(\boldsymbol{r})]$. In two spatial dimensions, the third term in Eq. \eqref{eq:hyd} can be absorbed into the pressure, so $\eta_\mathrm{o}$ does not contribute to $\boldsymbol{v}(\boldsymbol{r})$. It then follows from Eq. \eqref{eq:Bogdan} and the ordinary Stokes form for $\boldsymbol{\eta}^\mathrm{S}$ (first line in Eq. \eqref{eq:visctensor}), that $\eta_\mathrm{o}$ does not contribute to $\dot{E}$. This is a well-known result for this type of constitutive relation \citep{Banerjee:2017}. In contrast, in three (or higher) spatial dimensions, odd viscosity can contribute to the viscous dissipation. Here, we focus on the three-dimensional case.

In our analysis the fundamental solution to Eq. (\ref{eq:hyd}) is essential, which is defined as the response of $\boldsymbol{v}(\boldsymbol{r})$ and $\tilde{p}(\boldsymbol{r})$ to a point force density $\boldsymbol{f}(\boldsymbol{r})=\boldsymbol{F}_0\delta(\boldsymbol{r})$. This defines the Green's tensor $\mathsfbi{G}(\boldsymbol{r})$ and pressure vector $\boldsymbol{Q}(\boldsymbol{r})$ via $\boldsymbol{v}(\boldsymbol{r})=\mathsfbi{G}(\boldsymbol{r})\cdot\boldsymbol{F}_0$ and $\tilde{p}(\boldsymbol{r})=\boldsymbol{Q}(\boldsymbol{r})\cdot\boldsymbol{F}_0$. 
The Green's functions can be explicitly computed \citep{Everts:2024}. 
We define an orthonormal triad of spherical basis vectors $\{\boldsymbol{\hat{r}},\boldsymbol{\hat{\theta}},\boldsymbol{\hat{\phi}}\}$, with $\boldsymbol{\hat{\phi}}=(\boldsymbol{\hat{\ell}}\times\boldsymbol{\hat{r}})/|\boldsymbol{\hat{\ell}}\times\boldsymbol{\hat{r}}|$ and $\boldsymbol{\hat{\theta}}=\boldsymbol{\hat{\phi}}\times\boldsymbol{\hat{r}}$, and the auxiliary variable $s=\gamma|\boldsymbol{\hat{r}}\times\boldsymbol{\hat{\ell}}|$, where $\gamma=\eta_\mathrm{o}/\eta_\mathrm{s}$. With this notation, we find
\begin{gather}
\mathsfbi{G}(\boldsymbol{r})=\frac{\Lambda(s)}{4\pi\eta_\mathrm{s}r[1+\Lambda(s)]}\left[\mathsfbi{I}+\Lambda(s)\hat{\boldsymbol{r}}\hat{\boldsymbol{r}}+s\Lambda(s)(\boldsymbol{\hat{r}}\boldsymbol{\hat{\phi}}-\boldsymbol{\hat{\phi}}\boldsymbol{\hat{r}})
-\left[1-\Lambda(s)\right]\boldsymbol{\hat{\phi}}\boldsymbol{\hat{\phi}}\right], \label{eq:greenodd} \\
\boldsymbol{Q}(\boldsymbol{r})=\frac{1}{4\pi r^2}\boldsymbol{\hat{r}}, \label{eq:presvec}
\end{gather}
with $\Lambda(s)=(1+s^2)^{-1/2}$. Observe that Eq. \eqref{eq:presvec} is of the Stokes form and that  $\boldsymbol{\hat{r}}\boldsymbol{\hat{\phi}}-\boldsymbol{\hat{\phi}}\boldsymbol{\hat{r}}$ is the rotation operator around the $\boldsymbol{\hat{\theta}}$ axis. For the transformations $(\boldsymbol{r},\boldsymbol{\hat{\ell}})\rightarrow(\boldsymbol{r},-\boldsymbol{\hat{\ell}})$ and $(\boldsymbol{r},\boldsymbol{\hat{\ell}})\rightarrow(-\boldsymbol{r},\boldsymbol{\hat{\ell}})$, we have that $\boldsymbol{\hat{\phi}}\rightarrow -\boldsymbol{\hat{\phi}}$. We have the following symmetry properties
\begin{gather}
G_{\alpha\beta}(\boldsymbol{r};\boldsymbol{\hat{\ell}})=G_{\beta\alpha}(\boldsymbol{r};-\boldsymbol{\hat{\ell}}), \\
G_{\alpha\beta}(\boldsymbol{r};\boldsymbol{\hat{\ell}})=G_{\alpha\beta}(-\boldsymbol{r};\boldsymbol{\hat{\ell}}), \quad Q_\alpha(\boldsymbol{r};\boldsymbol{\hat{\ell}})=-Q_\alpha(-\boldsymbol{r};\boldsymbol{\hat{\ell}}). \label{eq:symmetry}
\end{gather}
The first line follows from the Onsager-Casimir relations, while the second line follows from the defining differential equation for $\mathsfbi{G}(\boldsymbol{r})$ and $\boldsymbol{Q}(\boldsymbol{r})$. These relations can also be deduced from the explicit expressions Eqs. \eqref{eq:greenodd} and \eqref{eq:presvec}.
We define the corresponding rate of strain tensor $\boldsymbol{\Theta}(\boldsymbol{r})$ and stress tensor $\boldsymbol{\Sigma}(\boldsymbol{r})$ through the relations $\mathsfbi{e}(\boldsymbol{r})=\boldsymbol{\Theta}(\boldsymbol{r})\cdot\boldsymbol{F}_0$ and $\boldsymbol{\sigma}(\boldsymbol{r})=\boldsymbol{\Sigma}(\boldsymbol{r})\cdot\boldsymbol{F}_0$. We find
\begin{equation}
\Sigma_{\alpha\beta\nu}(\boldsymbol{r})=-\delta_{\alpha\beta}P_{\nu}(\boldsymbol{r})+2\eta_\mathrm{s}\Theta_{\alpha\beta\nu}(\boldsymbol{r})+2\eta_\mathrm{o}\hat{\ell}_{\kappa}[\epsilon_{\kappa\alpha\lambda}\Theta_{\lambda\beta\nu}(\boldsymbol{r})+\epsilon_{\kappa\beta\lambda}\Theta_{\lambda\alpha\nu}(\boldsymbol{r})], \label{eq:fundstress}
\end{equation}
where
\begin{equation}
\boldsymbol{P}(\boldsymbol{r})=\boldsymbol{Q}(\boldsymbol{r})-2\eta_\mathrm{o}\boldsymbol{\hat{\ell}}\cdot[\nabla\times\mathsfbi{G}(\boldsymbol{r})] \label{eq:P}
\end{equation}
with $[\nabla\times\mathsfbi{G}(\boldsymbol{r})]_{\alpha\beta}=\epsilon_{\alpha\nu\lambda}\partial_{\nu} G_{\lambda\beta}(\boldsymbol{r})$. We have the symmetry relations 
\begin{equation}
    \Sigma_{\alpha\beta\nu}(\boldsymbol{r};\boldsymbol{\hat{\ell}})=\Sigma_{\beta\alpha\nu}(\boldsymbol{r};\boldsymbol{\hat{\ell}}), \quad \Sigma_{\alpha\beta\nu}(\boldsymbol{r};\boldsymbol{\hat{\ell}})=-\Sigma_{\alpha\beta\nu}(-\boldsymbol{r};\boldsymbol{\hat{\ell}}).
\end{equation}
Expressions for the elements of $\boldsymbol{\Sigma}(\boldsymbol{r})$ are listed in Appendix \ref{app} and are a crucial novelty of this work.

\section{Friction problem for a single sphere and its singularity representation}
Consider a sphere of radius $a$ in an unbounded fluid described by Eq. (\ref{eq:hyd}) for $r>a$ and $\boldsymbol{f}(\boldsymbol{r})=\boldsymbol{0}$. Now Eq. (\ref{eq:hyd}) is supplemented by the boundary conditions 
\begin{equation}
\boldsymbol{v}(a\hat{\boldsymbol{r}})=\boldsymbol{U}+\boldsymbol{\Omega}\times(a\hat{\boldsymbol{r}}), \qquad \boldsymbol{v}(\boldsymbol{r})\rightarrow \boldsymbol{v}_\infty(\boldsymbol{r}), \quad \tilde{p}(\boldsymbol{r})=\tilde{p}_\infty(\boldsymbol{r}), \quad(r\rightarrow\infty).
\end{equation}
Here, the solid-body motion of the sphere is characterized by translational velocity $\boldsymbol{U}$ and rotational velocity $\boldsymbol{\Omega}$, and we included an ambient flow and pressure field $\boldsymbol{v}_\infty(\boldsymbol{r})$ and $\tilde{p}_\infty(\boldsymbol{r})$, respectively.  
Our goal is to find the singularity representation for $\boldsymbol{v}(\boldsymbol{r})$ and $\tilde{p}(\boldsymbol{r})$, with corresponding stress tensor $\boldsymbol{\sigma}(\boldsymbol{r})$. To find this expression, we define an auxiliary flow field $\boldsymbol{v}_0(\boldsymbol{r})$ with stress tensor $\boldsymbol{\sigma}_0(\boldsymbol{r})$. Two such flows defined in a volume $\mathcal{V}$ bounded by a closed surface $\mathcal{S}$ are related by the Lorentz reciprocal theorem for odd fluids \citep{Vilfan:2023},
\begin{align}
&\int_{\mathcal{S}}\mathrm{d}S\, \boldsymbol{v}(\boldsymbol{r};\boldsymbol{\hat{\ell}})\cdot[\boldsymbol{\sigma}_0(\boldsymbol{r};-\boldsymbol{\hat{\ell}})\cdot\boldsymbol{\hat{n}}]-\int_\mathcal{V}\mathrm{d}V\, \boldsymbol{v}_0(\boldsymbol{r};-\boldsymbol{\hat{\ell}})\cdot[\nabla\cdot\boldsymbol{\sigma}(\boldsymbol{r};\boldsymbol{\hat{\ell}})] \nonumber\\ 
&=\int_{\mathcal{S}}\mathrm{d}S\, \boldsymbol{v}_0(\boldsymbol{r};-\boldsymbol{\hat{\ell}})\cdot[\boldsymbol{\sigma}(\boldsymbol{r};\boldsymbol{\hat{\ell}})\cdot\boldsymbol{\hat{n}}]-\int_\mathcal{V}\mathrm{d}V\, \boldsymbol{v}(\boldsymbol{r};\boldsymbol{\hat{\ell}})\cdot[\nabla\cdot\boldsymbol{\sigma}_0(\boldsymbol{r};-\boldsymbol{\hat{\ell}})].  
\end{align}
The form of this theorem is the same as in ordinary Stokes flow \citep{Lorentz, Masoud.Stone2019}, with the important difference that $\boldsymbol{v}_0$ satisfies Eq. \eqref{eq:hyd} with $\boldsymbol{\hat{\ell}}$ replaced by $-\boldsymbol{\hat{\ell}}$ (as indicated by the second argument). For the auxiliary flow field, we replace $\boldsymbol{v}_0(\boldsymbol{r};-\boldsymbol{\hat{\ell}})$ by $\mathsfbi{G}(\boldsymbol{r}'-\boldsymbol{r};-\boldsymbol{\hat{\ell}})$. Since we impose stick-boundary conditions, the double-layer contribution can be eliminated using the same steps as for ordinary Stokes flow, see Sec. 2.4.2 in \cite{Kim}. We find that the induced forces picture for a rigid body with surface $\mathcal{S}_\mathrm{p}$ in a chiral active fluid is identical to the one for ordinary Stokes flow \citep{Mazur:1974}
\begin{equation}
\boldsymbol{u}(\boldsymbol{r})-\boldsymbol{v}^\infty(\boldsymbol{r})=\int_\mathrm{\mathcal{S}_\mathrm{p}} \mathrm{d}S_{\boldsymbol{r}'}\, \mathsfbi{G}(\boldsymbol{r}-\boldsymbol{r}';\boldsymbol{\hat{\ell}})\cdot\boldsymbol{f}_\mathrm{ind}(\boldsymbol{r}'), \label{eq:ind}
\end{equation}
where $\boldsymbol{f}_\mathrm{ind}=\boldsymbol{\sigma}(\boldsymbol{r})\cdot\boldsymbol{\hat{n}}|_{\boldsymbol{r}\in\mathcal{S}_\mathrm{p}}$, with $\boldsymbol{\hat{n}}$ pointing towards the particle. Note that in \citep{Kim} $\mathsfbi{G}$ is located on the right side of $\boldsymbol{f}_\mathrm{ind}$. For the odd case, commuting it to the left side changes $-\boldsymbol{\hat{\ell}}$ to $+\boldsymbol{\hat{\ell}}$, using Eq. \eqref{eq:symmetry}. Eq. \eqref{eq:ind} is also valid inside the particle, where for a sphere $\boldsymbol{u}(\boldsymbol{r})=\boldsymbol{U}+\boldsymbol{\Omega}\times\boldsymbol{r}$ for $r<a$ and $\boldsymbol{u}(\boldsymbol{r})=\boldsymbol{v}(\boldsymbol{r})$ for $r>a$. Furthermore, $\boldsymbol{f}_\mathrm{ind}(\boldsymbol{r})$ has the first few moments given by
\begin{equation}
\boldsymbol{F}=\int_{\mathcal{S}_\mathrm{p}} \mathrm{d}S\, \boldsymbol{f}_\mathrm{ind}(\boldsymbol{r}), \quad
\boldsymbol{T}= \int_{\mathcal{S}_\mathrm{p}} \mathrm{d}S\, \boldsymbol{r}\times\boldsymbol{f}_\mathrm{ind}(\boldsymbol{r}), \quad
\mathsfbi{S}= \int_{\mathcal{S}_\mathrm{p}} \mathrm{d}S\, \overbracket{{\boldsymbol{r}\, \boldsymbol{f}_\mathrm{ind}(\boldsymbol{r})}}, \label{eq:forces}
\end{equation}
with $\boldsymbol{F}$ the force, $\boldsymbol{T}$ the torque, and $\mathsfbi{S}$ the stresslet of the particle acting on the fluid. The overbracket  defines the symmetric traceless part of a tensor. For later use, we also define the grand friction tensor components as,
\begin{equation}
\begin{pmatrix}
\boldsymbol{F}\\
\boldsymbol{T}\\
\mathsfbi{S}
\end{pmatrix}
=
\begin{pmatrix}
\boldsymbol{\zeta}^\mathrm{tt} & \boldsymbol{\zeta}^\mathrm{tr} & \boldsymbol{\zeta}^\mathrm{td}\\
\boldsymbol{\zeta}^\mathrm{rt} & \boldsymbol{\zeta}^\mathrm{rr} & \boldsymbol{\zeta}^\mathrm{rd}\\
\boldsymbol{\zeta}^\mathrm{dt} & \boldsymbol{\zeta}^\mathrm{dr} & \boldsymbol{\zeta}^\mathrm{dd}
\end{pmatrix}
\begin{pmatrix}
\boldsymbol{U}-\boldsymbol{U}^\infty\\
\boldsymbol{\Omega}-\boldsymbol{\Omega}^\infty\\
-\mathsfbi{E}^\infty,
\end{pmatrix},
\end{equation}
where the first moments of $\boldsymbol{v}^\infty(\boldsymbol{r})$ give a contribution from constant ambient flow $\boldsymbol{U}^\infty$, a rotating ambient velocity $\boldsymbol{\Omega}^\infty$, and a linear straining field $\mathsfbi{E}^\infty$.

For a sphere, Eq. (\ref{eq:ind}) simplifies to 
\begin{equation}
\boldsymbol{u}(\boldsymbol{r})-\boldsymbol{v}^\infty(\boldsymbol{r})=a^2\int_{S^2} \mathrm{d}^2\hat{\boldsymbol{r}}' \, \mathsfbi{G}(\boldsymbol{r}-a\hat{\boldsymbol{r}}')\cdot\boldsymbol{{f}}_\mathrm{ind}(\hat{\boldsymbol{r}}'), \label{eq:ind2}
\end{equation}
with $S^2$ the two-dimensional unit sphere.
For the case $r\geq a$, we take the sources in the centre and perform a multipole expansion
\begin{equation}
\mathsfbi{G}(\boldsymbol{r}-a\hat{\boldsymbol{r}}')=\left[1-a\hat{\boldsymbol{r}}'\cdot\frac{\partial}{\partial\boldsymbol{r}}+\frac{a^2}{2}\left(\hat{\boldsymbol{r}}'\cdot\frac{\partial}{\partial\boldsymbol{r}}\right)^2+...\right]\mathsfbi{G}(\boldsymbol{r})=:\mathrm{e}^{-a\hat{\boldsymbol{r}}'\cdot\nabla}\mathsfbi{G}(\boldsymbol{r}), 
\end{equation}
which results in
\begin{subeqnarray}
\boldsymbol{v}(\boldsymbol{r})-\boldsymbol{v}^\infty(\boldsymbol{r})&=&a^2\int_{S^2} \mathrm{d}^2\hat{\boldsymbol{r}}'\, \mathrm{e}^{-a\hat{\boldsymbol{r}}'\cdot\nabla}\mathsfbi{G}(\boldsymbol{r})\cdot\boldsymbol{{f}}_\mathrm{ind}(\hat{\boldsymbol{r}}'), \\ \label{eq:ind3v}
\tilde{p}(\boldsymbol{r})-\tilde{p}^\infty(\boldsymbol{r})&=&a^2\int_{S^2} \mathrm{d}^2\hat{\boldsymbol{r}}'\, \mathrm{e}^{-a\hat{\boldsymbol{r}}'\cdot\nabla}\boldsymbol{Q}(\boldsymbol{r})\cdot\boldsymbol{{f}}_\mathrm{ind}(\hat{\boldsymbol{r}}'). \label{eq:ind3p}
\end{subeqnarray}
To find the singularity representation of $\tilde{p}(\boldsymbol{r})$ we used  that $\boldsymbol{Q}(\boldsymbol{r})$ is of the same form as for a Newtonian fluid. We note that for more complicated boundary conditions (e.g., droplets immersed in an odd fluid), Eq. \eqref{eq:ind3v} contains not only the single-layer contributions, but also terms from the hydrodynamic double layer. Such contributions can be computed using the results in Appendix \ref{app}.  

\section{Translating sphere in constant ambient flow}
\label{sec:trans}
\subsection{Singularity representation for a translating sphere}

We first consider Eq. \eqref{eq:ind3v} for the case with $\boldsymbol{\Omega}=0$ and a constant ambient flow field $\boldsymbol{v}^\infty(\boldsymbol{r})=\boldsymbol{U}^\infty$. As an ansatz, we assume that this case is described by a constant $\boldsymbol{{f}}_\mathrm{ind}(\boldsymbol{\hat{r}})$. Using Eq. (\ref{eq:forces}) we find then that $\boldsymbol{F}=4\pi a^2\boldsymbol{{f}}_\mathrm{ind}$ and Eq. (\ref{eq:ind3v}{\it{a}}) reduces to
\begin{equation}
\boldsymbol{v}(\boldsymbol{r})-\boldsymbol{U}^\infty=\frac{1}{4\pi}\int_{S^2} \mathrm{d}^2\hat{\boldsymbol{r}}'\, \mathrm{e}^{-a\hat{\boldsymbol{r}}'\cdot\nabla}\mathsfbi{G}(\boldsymbol{r})\cdot\boldsymbol{F}=:\mathcal{L}_0\mathsfbi{G}(\boldsymbol{r})\cdot\boldsymbol{F}. \label{eq:singU}
\end{equation}
Note that within this ansatz, it follows from Eq. (\ref{eq:forces}) that $\boldsymbol{T}=0$ and $\mathsfbi{S}=0$ and, therefore, there is no translational-rotational and translational-dipolar coupling (i.e., $\boldsymbol{\zeta}^\mathrm{tr}=\boldsymbol{\zeta}^\mathrm{rt}=0$ and $\boldsymbol{\zeta}^\mathrm{td}=\boldsymbol{\zeta}^\mathrm{dt}=0$). Thus we can write
\begin{equation}
\boldsymbol{v}(\boldsymbol{r})-\boldsymbol{U}^\infty=\mathcal{L}_0\mathsfbi{G}(\boldsymbol{r})\cdot\boldsymbol{\zeta}^\mathrm{tt}\cdot(\boldsymbol{U}-\boldsymbol{U}^\infty), \label{eq:bla}
\end{equation}
where $\boldsymbol{\zeta}^\mathrm{tt}$ is the translational-translational friction tensor.
The $\mathcal{L}_0$ operator can be explicitly found by the angular integration in Eq. (\ref{eq:singU}), 
\begin{equation}
\mathcal{L}_0=\frac{1}{4\pi}\int_{S^2} \mathrm{d}^2\hat{\boldsymbol{r}}'\, \mathrm{e}^{-a\hat{\boldsymbol{r}}'\cdot\nabla}=\sum_{n=0}^\infty\frac{a^{2n}}{(2n+1)!}(\nabla^2)^n=:j_0(\mathrm{i}\mathcal{D}), \label{eq:lnot}
\end{equation}
with $\mathcal{D}^2=a^2\nabla^2$ and $j_n$ the $n$th order spherical Bessel function of the first kind. 

The operator series acting on $\mathsfbi{G}(\boldsymbol{r})$ can be evaluated using Fourier methods \citep{Everts:2024}
\begin{equation}
\mathcal{L}_0\mathsfbi{G}(\boldsymbol{r})=\frac{1}{\eta_\mathrm{s}}\int\frac{\mathrm{d}^3\boldsymbol{k}}{(2\pi)^3}\, j_0(ka)\mathrm{e}^{\mathrm{i}\boldsymbol{k}\cdot\boldsymbol{r}}\,\frac{\mathsfbi{I}-\hat{\boldsymbol{k}}\hat{\boldsymbol{k}}+\gamma (\hat{\boldsymbol{k}}\cdot\boldsymbol{\hat{\ell}})(\boldsymbol{\epsilon}\cdot\hat{\boldsymbol{k}})}{k^2[1+\gamma^2(\hat{\boldsymbol{k}}\cdot\boldsymbol{\hat{\ell}})^2]}. \label{eq:Fourierrep}
\end{equation}
To show that the initial ansatz of a constant $\boldsymbol{{f}}_\mathrm{ind}$ is correct, we observe that Eq. (\ref{eq:singU}) satisfies Eq. (\ref{eq:hyd}). We thus need to check only whether the boundary conditions are satisfied. Indeed, in \cite{Everts:2024} it was shown that $\mathcal{L}_0\mathsfbi{G}(a\hat{\boldsymbol{r}})$ is independent of $\boldsymbol{\hat{r}}$ and, therefore, the boundary condition $\boldsymbol{v}(a\hat{\boldsymbol{r}})=\boldsymbol{U}$ can be satisfied when we identify $\boldsymbol{\zeta}^\mathrm{tt}=[\mathcal{L}_0\mathsfbi{G}(a\hat{\boldsymbol{r}})]^{-1}$. We conclude that Eq. (\ref{eq:bla}) solves the boundary-value problem and that $\boldsymbol{{f}}_\mathrm{ind}$ is indeed constant. Explicit evaluation of $\boldsymbol{\zeta}^\mathrm{tt}$  gives
\begin{equation}
\boldsymbol{\zeta}^\mathrm{tt}=24\pi\eta_\mathrm{s}a\Bigg[\frac{R(\gamma)(\mathsfbi{I}-\boldsymbol{\hat{\ell}\hat{\ell}})+S(\gamma)(\boldsymbol{\epsilon}\cdot\boldsymbol{\hat{\ell}})}{R(\gamma)^2+S(\gamma)^2}+\frac{\boldsymbol{\hat{\ell}\hat{\ell}}}{T(\gamma)}\Bigg]. \label{eq:zetatt}
\end{equation}
Here, we defined the following functions of the parameter $\gamma=\eta_\mathrm{o}/\eta_\mathrm{s}$
\begin{gather}
R(\gamma)=\gamma^2[f(\gamma)-g(\gamma)]+4, \quad S(\gamma)=2\gamma f(\gamma),\quad
T(\gamma)=2\gamma^2[f(\gamma)+g(\gamma)]+4, 
\end{gather}
with 
\begin{equation}
f(\gamma)=\frac{3}{\gamma^2}\left(\frac{\arctan{\gamma}}{\gamma}-1\right)=-1+O(\gamma^2),\quad {g(\gamma)=\frac{1+f(\gamma)}{\gamma^2}=\frac{3}{5}+O(\gamma^2)}. 
\end{equation}
Eqs. (\ref{eq:bla}) and (\ref{eq:zetatt}) form the singularity solution for a translating sphere. The first-order approximate result for $\gamma\ll 1$ found by expanding Eq. (\ref{eq:zetatt}) is consistent with results found in \cite{Khain:2022} and \cite{Cruz:2023}. Furthermore, since $(\nabla^2)^n\boldsymbol{Q}(\boldsymbol{r})=0$ for $n\geq2$, we find for the effective pressure using  Eq. (\ref{eq:ind3v}{\it{b}})
\begin{equation}
\tilde{p}(\boldsymbol{r})-\tilde{p}^\infty=\boldsymbol{Q}(\boldsymbol{r})\cdot\boldsymbol{\zeta}^\mathrm{tt}\cdot(\boldsymbol{U}-\boldsymbol{U}^\infty), \label{eq:effpres}
\end{equation}
with $\tilde{p}^\infty$ a constant.
Note that for $\gamma=0$, we find the Oseen tensor $\mathsfbi{G}(\boldsymbol{r})|_{\gamma=0}=(8\pi\eta_\mathrm{s}r)^{-1}(\mathsfbi{I}+\hat{\boldsymbol{r}}\hat{\boldsymbol{r}})$ and since $(\nabla^2)^n\mathsfbi{G}(\boldsymbol{r})|_{\gamma=0}=0$ for $n\geq 2$, we find the well-known singularity representations for ordinary Stokes flow
\begin{subeqnarray}
\boldsymbol{v}(\boldsymbol{r})|_{\gamma=0}-\boldsymbol{U}^\infty&=&6\pi\eta_\mathrm{s}a\left(1+\frac{a^2}{6}\nabla^2\right)\mathsfbi{G}(\boldsymbol{r})|_{\gamma=0}\cdot(\boldsymbol{U-U}^\infty), \\
p(\boldsymbol{r})|_{\gamma=0}-p^\infty&=&\frac{3a}{2\eta_\mathrm{s} r^2}\hat{\boldsymbol{r}}\cdot(\boldsymbol{ U-U}^\infty).
\end{subeqnarray}

\subsection{Explicit evaluation of the fluid velocity field around a translating sphere}
Now we consider $\boldsymbol{v}(\boldsymbol{r})$ for $\gamma\neq 0$. Since the singularity representation Eq. (\ref{eq:singU}) solves the boundary value problem of a translating sphere, we can obtain an explicit form for $\boldsymbol{v}(\boldsymbol{r})$ by evaluating (\ref{eq:Fourierrep}) for all $r>a$. The details are presented in Appendix \ref{appA}.
The final result is 
\begin{align}
\boldsymbol{v}(\boldsymbol{r})=&\frac{6}{\gamma^2T(\gamma)}\left\{\left[\frac{1+\gamma^2}{\gamma}\mathcal{M}(\boldsymbol{r};\gamma)-\frac{a}{r}\right]\boldsymbol{\hat{\ell}}\boldsymbol{\hat{\ell}}+\mathcal{N}(\boldsymbol{r};\gamma)(\gamma\boldsymbol{\hat{\phi}}\boldsymbol{\hat{\ell}}-\boldsymbol{\hat{\rho}}\boldsymbol{\hat{\ell}})\right\}\cdot(\boldsymbol{U}-\boldsymbol{U}^\infty) \nonumber \\
&+\frac{6}{\gamma^2[R(\gamma)^2+S(\gamma)^2]}\Bigg(\mathcal{O}(\boldsymbol{r};\gamma)\left[R(\gamma)\boldsymbol{\hat{\rho}}\boldsymbol{\hat{\rho}}+S(\gamma)\boldsymbol{\hat{\rho}}\boldsymbol{\hat{\phi}}\right]\nonumber\\
&+\mathcal{N}(\boldsymbol{r};\gamma)\left\{[\gamma S(\gamma)-R(\gamma)]\boldsymbol{\hat{\ell}}\boldsymbol{\hat{\rho}}-[\gamma R(\gamma)+S(\gamma)]\boldsymbol{\hat{\ell}}\boldsymbol{\hat{\phi}}\right\} \label{eq:v}\\
&-\left[\frac{1}{2}\mathcal{O}(\boldsymbol{r};\gamma)+\frac{1-\gamma^2}{2\gamma}\mathcal{M}(\boldsymbol{r};\gamma)-\frac{a}{2 r}\right]\left[R(\gamma)(\mathsfbi{I}-\boldsymbol{\hat{\ell}\hat{\ell}})+S(\gamma)(\boldsymbol{\epsilon}\cdot\boldsymbol{\hat{\ell}})\right]\nonumber\\
&-\left[\mathcal{M}(\boldsymbol{r};\gamma)-\frac{a\gamma}{r}\right]\left[R(\gamma)(\boldsymbol{\epsilon}\cdot\boldsymbol{\hat{\ell}})-S(\gamma)(\mathsfbi{I}-\boldsymbol{\hat{\ell}\hat{\ell}})\right]
\Bigg)\cdot(\boldsymbol{U}-\boldsymbol{U}^\infty) \nonumber
\end{align}
with the dimensionless functions 
\begin{gather}
\mathcal{M}(\boldsymbol{r};\gamma)=\mathrm{arcsin}\left[\frac{1}{\mathcal{R}_+(\boldsymbol{r};\gamma)}\right],\nonumber \\
    \mathcal{N}(\boldsymbol{r};\gamma)=\frac{a}{r|\boldsymbol{\hat{r}}\times\boldsymbol{\hat{\ell}}|}\left[\mathrm{sgn}(\boldsymbol{\hat{r}}\cdot\boldsymbol{\hat{\ell}})\sqrt{1-\mathcal{R}_-(\boldsymbol{r};\gamma)^2}-(\boldsymbol{\hat{r}}\cdot\boldsymbol{\hat{\ell}})\right],\label{eq:something}  \\
    \mathcal{O}(\boldsymbol{r};\gamma)=\frac{a^2}{r^2|\boldsymbol{\hat{r}}\times\boldsymbol{\hat{\ell}}|^2}\left\{\gamma\sqrt{\mathcal{R}_+(\boldsymbol{r};\gamma)^2-1}\left[1-\sqrt{1-\mathcal{R}_-(\boldsymbol{r};\gamma)^2}\right]^2-\frac{r}{ a}(1-|\boldsymbol{\hat{r}}\cdot\boldsymbol{\hat{\ell}}|)^2\right\} \nonumber 
\end{gather} defined in terms of
\begin{gather}
\mathcal{R}_\pm(\boldsymbol{r};\gamma)=\frac{r}{2a}\frac{\mathcal{A}_+\pm\mathcal{A}_-}{\sin\psi}, \quad \mathcal{A}_\pm(\boldsymbol{r};\gamma)=\sqrt{(\boldsymbol{\hat{r}}\cdot\boldsymbol{\hat{\ell}})^2\cos^2\psi+\left(|\boldsymbol{\hat{r}}\times\boldsymbol{\hat{\ell}}|\pm \frac{a}{r}\sin\psi\right)^2}, \label{eq:rpm} 
\end{gather}
and $\psi=\mathrm{arctan}(\gamma)$. Moreover, we have adopted a cylindrical basis $\{\boldsymbol{\hat{\rho}},\boldsymbol{\hat{\phi}},\boldsymbol{\hat{\ell}}\}$ with $\boldsymbol{\hat{\rho}}=\boldsymbol{\hat{\phi}}\times\boldsymbol{\hat{\ell}}$. Note that $\boldsymbol{\hat{\rho}}\rightarrow\boldsymbol{\hat{\rho}}$ under the transformation $\boldsymbol{\hat{\ell}}\rightarrow-\boldsymbol{\hat{\ell}}$. The first line in Eq. \eqref{eq:v} corresponds to the flow field for $\boldsymbol{U}\parallel\boldsymbol{\hat\ell}$, which was already found in previous work \citep{Everts:2024}. However, in this work, the closed-form expressions for the case $\boldsymbol{U}\perp\boldsymbol{\hat\ell}$ were not presented, although the solution was sketched in the Supplemental Material. Here, we have simplified this calculation by rewriting certain integrals using identities of Bessel functions and explicitly carried out this computation until the end. Our new result Eq. \eqref{eq:v} is valid for arbitrary directions of $\boldsymbol{U}$ and $\boldsymbol{\hat\ell}$ and it is worth noting that it is an analytical, exact closed-form expression for a fully three-dimensional flow field around a sphere in a chiral active fluid. Furthermore, Eq. \eqref{eq:v} can be expanded for small $\gamma$ and $\boldsymbol{U}^\infty=\boldsymbol{0}$ as
\begin{align}
\boldsymbol{v}(\boldsymbol{r})=&\left[\frac{3a}{4r}\left(\mathsfbi{I}+\boldsymbol{\hat r}\boldsymbol{\hat r}\right)+\frac{a^3}{4r^3}\left(\mathsfbi{I}-3\boldsymbol{\hat r}\boldsymbol{\hat r}\right)\right]\cdot\boldsymbol{U}\nonumber\\-&\frac{3a}{8r}\left(1-\frac{a^2}{r^2}\right)\gamma\left[\boldsymbol{\hat\ell}\times\boldsymbol{U}-2(\boldsymbol{\hat r}\times\boldsymbol{U})(\boldsymbol{\hat r}\cdot\boldsymbol{\hat\ell})-\boldsymbol{\hat r}\cdot(\boldsymbol{\hat\ell}\times\boldsymbol{U})\boldsymbol{\hat r}\right]+O(\gamma^2).
\end{align} The first term corresponds to the ordinary Stokes solution of a Newtonian fluid \citep{Kim}, and the second term corresponds to the solution first calculated by \citet{Khain:2022}.
\begin{figure}
    \centering
    \includegraphics[width=1\linewidth]{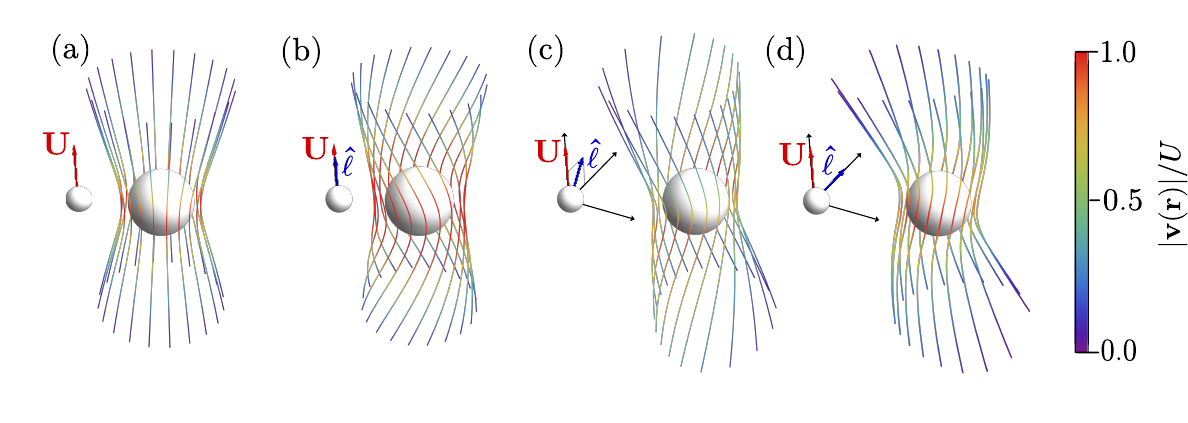}
    \caption{Representative streamlines of the fluid velocity field $\boldsymbol{v}(\boldsymbol{r})$ around a spherical particle translating with velocity $\boldsymbol{U}$ in the absence of ambient flow scaled to $U=|\boldsymbol{U}|$. All plots are generated using the exact analytical solution Eq. \eqref{eq:v}. (a) Stokes flow without odd viscosity ($\gamma =\eta_\mathrm{o}/\eta_\mathrm{s}=0$). (b)-(d) Odd viscous flow for $\gamma =3$ at different relative orientations of $\boldsymbol{U}$ and $\boldsymbol{\hat\ell}$, with (b) $\boldsymbol{U}\parallel\boldsymbol{\hat\ell}$, (c) $\sphericalangle(\boldsymbol{U},\boldsymbol{\hat\ell})=45^\circ$, and (d) $\boldsymbol{U}\perp\boldsymbol{\hat\ell}.$ Note that the flows in (a) and (b) are cylindrically symmetric around the $\boldsymbol{U}$ axis. All streamlines are directed from bottom to top.}
    \label{fig:3dplots}
\end{figure}

\begin{figure}[ht!]
    \centering
    \includegraphics[width=1\linewidth]{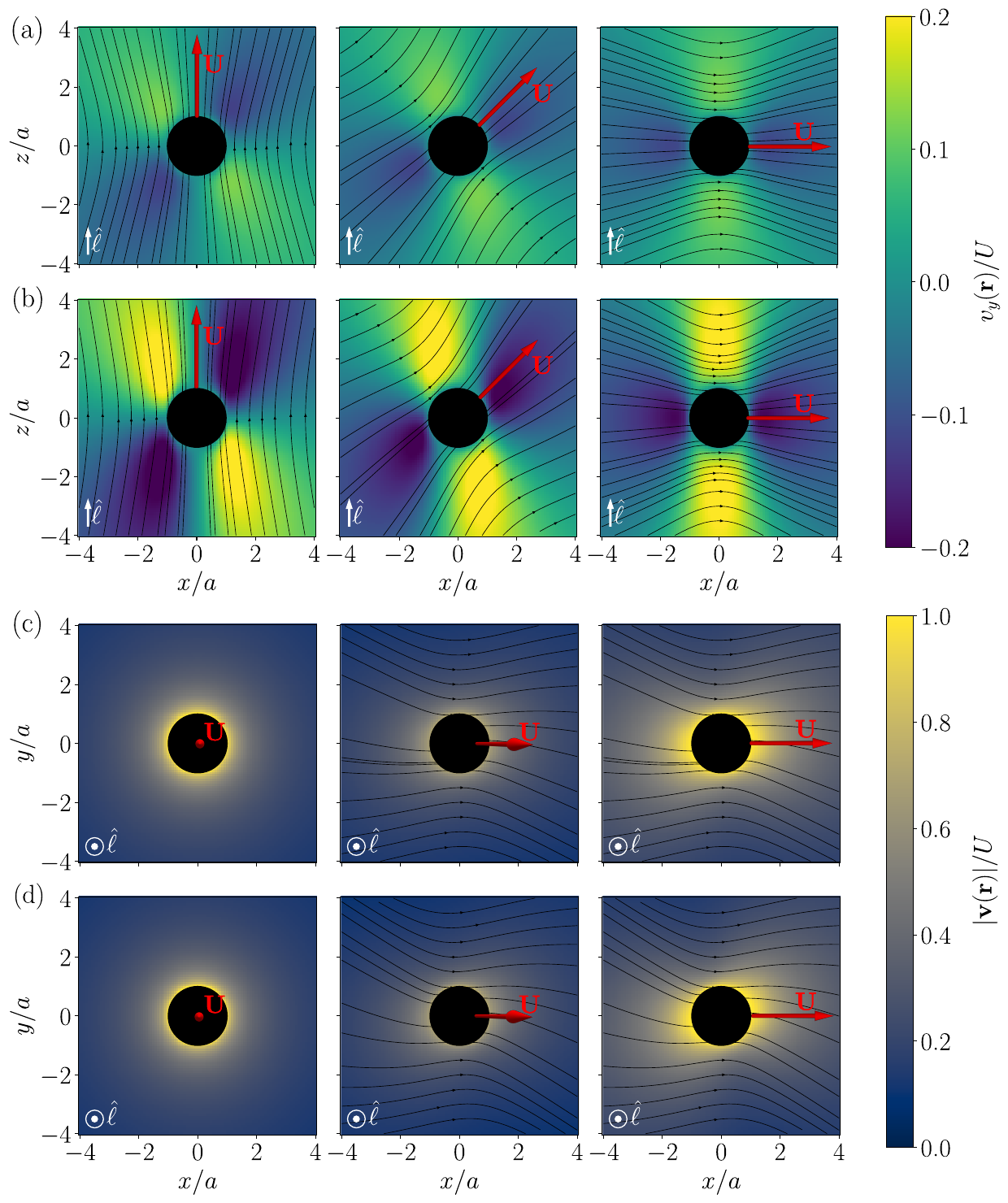}
    \caption{Projections of $\boldsymbol{v}(\boldsymbol{r})$ around a spherical particle translating with velocity $\boldsymbol{U}$ in an odd viscous fluid. The anisotropy axis $\boldsymbol{\hat\ell}$ is fixed in the $+z$ direction with the orientation of $\boldsymbol{U}$ being varied with respect to $\boldsymbol{\hat\ell}$. (a,b) Projections of the streamlines of $\boldsymbol{v}(\boldsymbol{r})$ onto the $xz$-plane. The colours are a measure for the out-of-plane component of $\boldsymbol{v}(\boldsymbol{r})$ in the $y$ direction scaled to $U=|\boldsymbol{U}|$. (c,d) Projections of the streamlines of $\boldsymbol{v}(\boldsymbol{r})$ onto the $xy$-plane, with the colours being a measure for $|\boldsymbol{v}(\boldsymbol{r})|$. Panels (a) and (c) are for $\gamma=1$, whereas panels (b) and (d) are for $\gamma=3$.}
    \label{fig:flow_around_sphere}
\end{figure}
Evaluating Eq. \eqref{eq:effpres}, we find for the effective pressure
\begin{equation}
\tilde{p}(\boldsymbol{r})-\tilde{p}^\infty=\frac{6\eta_\mathrm{s}a}{r^2}\Bigg\{\frac{|\boldsymbol{\hat{r}}\times\boldsymbol{\hat{\ell}}|\left[R(\gamma)\boldsymbol{\hat{\rho}}+S(\gamma)\boldsymbol{\hat{\phi}}\right] }{R(\gamma)^2+S(\gamma)^2}+\frac{(\boldsymbol{\hat{r}}\cdot\boldsymbol{\hat{\ell}})\boldsymbol{\hat{\ell}}}{T(\gamma)}\Bigg\}\cdot(\boldsymbol{U}-\boldsymbol{U}^\infty)
\end{equation}
Since the velocity field is affected by $\eta_\mathrm{o}$, it follows from Eq. \eqref{eq:Bogdan} that a translating sphere dissipates more energy in the odd model fluid than in a Newtonian fluid. The same result was obtained by \citet{Everts:2024} by explicitly computing $\dot{E}=\boldsymbol{\zeta}^\mathrm{tt}:(\boldsymbol{U}-\boldsymbol{U}^\infty)(\boldsymbol{U}-\boldsymbol{U}^\infty)$ which is larger than the Stokesian dissipation $6\pi\eta_\mathrm{s}a|\boldsymbol{U}-\boldsymbol{U}^\infty|^2$ for all $\gamma>0$ and directions of $\boldsymbol{U}-\boldsymbol{U}^\infty$ with respect to $\boldsymbol{\hat{\ell}}$.

Using Eq. \eqref{eq:v}, we visualise the streamlines of $\boldsymbol{v}(\boldsymbol{r})$ for $\gamma=3$ in Fig.  \ref{fig:3dplots}(b)-(d) and compare it with ordinary Stokes flow $(\gamma=0)$, see Fig.  \ref{fig:3dplots}(a). The details of the flow and the effect of changing $\gamma$ can be seen in projections onto two orthogonal planes in Fig. \ref{fig:flow_around_sphere}(a)-(d). A distinct feature of odd viscosity is the emergence of azimuthal flows, which are most pronounced for $\boldsymbol{U}\parallel\boldsymbol{\hat{\ell}}$ (Fig. \ref{fig:3dplots}(b)) where the flow is cylindrically symmetric. These azimuthal patterns can still be seen when $\boldsymbol{U}$ is not parallel to $\boldsymbol{\hat\ell}$, but with a tilted axis, as shown in Fig. \ref{fig:3dplots}(c,d) and in the projections Fig. \ref{fig:flow_around_sphere}(a,b).  Such patterns were first seen for a point-force response parallel to $\boldsymbol{\hat{\ell}}$, see \cite{Khain:2022}. The flows for a general direction of the point force follow from Eq. \eqref{eq:greenodd}.

\section{Rotating sphere}
\label{sec:rotsphere}
\subsection{Solution for a rotating sphere without using the singularity representation}
\label{sec:simple}
Consider now a rotating sphere without translation ($\boldsymbol{U}=\boldsymbol{0}$) in an ambient flow field $\boldsymbol{v}^\infty(\boldsymbol{r})=\boldsymbol{\Omega}^\infty\times\boldsymbol{r}$.
Before evaluating the singularity representation for this case, we note that there is an alternative solution method \citep{Hosaka:2024} by transforming Eq. \eqref{eq:hyd} to
\begin{equation}
\eta_\mathrm{s}\nabla^2\boldsymbol{ v}(\boldsymbol{r})-\nabla\hat p(\boldsymbol{r})-\eta_\mathrm{o}\nabla^2[\boldsymbol{ v}(\boldsymbol{r})\times\boldsymbol{\hat{\ell}}]=\boldsymbol{0}, \quad \nabla\cdot\boldsymbol{ v}(\boldsymbol{r})=0 \label{eq:hyd2}
\end{equation}
with modified effective pressure \citep{Cruz:2023}
\begin{equation}
    \hat{p}(\boldsymbol{r})=\tilde{p}(\boldsymbol{r})-\eta_\mathrm{o}\boldsymbol{\hat{\ell}}\cdot[\nabla\times\boldsymbol{v}(\boldsymbol{r})].
\end{equation}
We make the following important observation. The general solution for ordinary Stokes flow ($\eta_\mathrm{o}=0$) is given by the Lamb solution \citep{Lamb:1924}, which can be classified according to the irreducible representations of the rotational group by the multipoles $\boldsymbol{v}_{lm\sigma}^-(\boldsymbol{r})$ (singular at the origin) and $\boldsymbol{v}_{lm\sigma}^+(\boldsymbol{r})$ (singular at infinity), with $l=1,2,..$, $m=-l,...,l$, and $\sigma=0,1,2$. See \cite{Cichocki:1988} for explicit expressions of $\boldsymbol{v}_{lm\sigma}^\pm(\boldsymbol{r})$. We find that
\begin{subeqnarray}    
\nabla^2 \boldsymbol{v}_{lm\sigma}^-(\boldsymbol{r})=\boldsymbol{0}, \quad \nabla\cdot\boldsymbol{v}_{lm\sigma}^-(\boldsymbol{r})=0, \quad \mathrm{for}\ \sigma=1,2, \\
\nabla^2 \boldsymbol{v}_{lm\sigma}^+(\boldsymbol{r})=\boldsymbol{0}, \quad \nabla\cdot\boldsymbol{v}_{lm\sigma}^+(\boldsymbol{r})=0, \quad \mathrm{for}\ \sigma=0,1, \label{eq:multipole}
\end{subeqnarray}
i.e., they are the general solutions to the ordinary Stokes equations with constant pressure. Clearly, these multipoles are also solutions to Eq. \eqref{eq:hyd2} with $\hat{p}(\boldsymbol{r})$ constant.  Therefore, Eq. \eqref{eq:multipole} constitutes all solutions from the ordinary Stokes equations that are also solutions to the odd Stokes equations Eq. \eqref{eq:hyd} with $\boldsymbol{f=0}$. If supported by the appropriate boundary conditions, it follows from our generalised Helmholtz theorem Sec. \ref{sec:diss} that such flows have no contribution from odd viscosity to the viscous dissipation. The solution by \cite{Hosaka:2024} is an explicit example of this. Namely, the $\boldsymbol{v}(\boldsymbol{r})$ that satisfies the boundary condition of a rotating sphere can be constructed from $\boldsymbol{v}_{1m1}^-(\boldsymbol{r})$, to give the rotlet solution $\boldsymbol{v}(\boldsymbol{r})={-a^3(\boldsymbol{\Omega}-\boldsymbol{\Omega}^\infty)\times\nabla\left(1/r\right)}$ -- the same as in an ordinary Stokes flow \citep{Kim}. Since $\mathsfbi{E}^\infty=0$, we have $\boldsymbol{T}=\boldsymbol{\zeta}^\mathrm{rr}\cdot(\boldsymbol{\Omega}-\boldsymbol{\Omega}^\infty)$ and $\mathsfbi{S}=\boldsymbol{\zeta}^\mathrm{dr}\cdot(\boldsymbol{\Omega}-\boldsymbol{\Omega}^\infty)$. It follows from the corresponding stress tensor that 
\begin{equation}
\boldsymbol{\zeta}^\mathrm{rr}={8\pi\eta_\mathrm{s}a^3}\left[\mathsfbi{I}-\frac{\gamma}{2}(\boldsymbol{\epsilon}\cdot\boldsymbol{\hat{\ell}}) \label{eq:frictionrot}
\right], \quad \boldsymbol{\zeta}^\mathrm{dr}=4\pi\eta_\mathrm{s}a^3\gamma \mathsfbi{s},
\end{equation}
where $s_{\alpha\beta\nu}=(\hat{\ell}_{\alpha}\delta_{\beta\nu}+\hat{\ell}_{\beta}\delta_{\alpha\nu})/2-(1/3)\delta_{\alpha\beta}\hat{\ell}_\nu$. This is different from Newtonian fluids, where rotation of a suspended solid body does not produce a stresslet. Furthermore, from symmetry we also find $\boldsymbol{\zeta}^\mathrm{rd}$, since ${\zeta}_{\alpha\beta\lambda}^\mathrm{rd}(\boldsymbol{\hat{\ell}})={\zeta}_{\lambda\alpha\beta}^\mathrm{dr}(-\boldsymbol{\hat{\ell}})$. Taking a reference fluid with $\eta_\mathrm{o}=0$, we indeed conclude from Eq. \eqref{eq:Bogdan} that the viscous dissipation is the same as in ordinary Stokes flow. The same conclusion is obtained from $\dot{E}=\boldsymbol{\zeta}^\mathrm{rr}:(\boldsymbol{\Omega}-\boldsymbol{\Omega}^\infty)(\boldsymbol{\Omega}-\boldsymbol{\Omega}^\infty)$ with the $\boldsymbol{\zeta}^\mathrm{rr}$ from Eq. \eqref{eq:frictionrot}. 

Although this solution strategy is elegant, we cannot find $\boldsymbol{\zeta}^\mathrm{dd}$ nor can we describe $\boldsymbol{v}(\boldsymbol{r})$ in the presence of a linear ambient shear flow using this method. Furthermore, if multiple odd viscosities \citep{Khain:2022} are present, then this method breaks down as well. Therefore, it is important to find the general singularity representation for this type of boundary value problem.

\subsection{Solution from singularity representation and rotational dissipation}
To construct the singularity representation for the rotating sphere, we assume an induced force density that is linear in $\hat{\boldsymbol{r}}$, i.e. $\boldsymbol{{f}}_\mathrm{ind}(\hat{\boldsymbol{r}})=\mathsfbi{K}\cdot\hat{\boldsymbol{r}}$, with a constant tensor $\mathsfbi{K}$. Generally, $\mathsfbi{K}$ can be decomposed as
\begin{equation}
K_{\alpha\beta}=K^{(1)}\delta_{\alpha\beta}+K^{(2)}_\lambda\epsilon_{\lambda\alpha\beta}+K_{\alpha\beta}^{(3)},
\end{equation}
with a constant scalar $K^{(1)}$, vector $\boldsymbol{K}^{(2)}$, and symmetric traceless tensor $\mathsfbi{K}^{(3)}$. These quantities are determined below. Without loss of generality, we can set $K^{(1)}=0$ due to the incompressibility condition. Using Eq. (\ref{eq:forces}), we find $\boldsymbol{F}=\boldsymbol{0}$, $\boldsymbol{T}=-(8/3)\pi a ^3\boldsymbol{K}^{(2)}$, and $\mathsfbi{S}=(4/3)\pi a^3\mathsfbi{K}^{(3)}$. Then Eq. \eqref{eq:ind3v} reduces to
\begin{equation}
[\boldsymbol{v}(\boldsymbol{r})-\boldsymbol{\Omega}^\infty\times\boldsymbol{r}]_\alpha=-[\mathcal{L}_1\partial_\nu G_{\alpha\beta}](\boldsymbol{r})\left(\frac{1}{2}\epsilon_{\nu\beta\lambda}T_\lambda+S_{\nu\beta}\right),
\end{equation}
with the formal operator expression
\begin{equation}
\mathcal{L}_1\nabla=-\frac{3}{4\pi a}\int_{S^2}\mathrm{d}^2\boldsymbol{\hat{r}}'\, \boldsymbol{\hat{r}}' \mathrm{e}^{-a\boldsymbol{\hat{r}}'\cdot\nabla}.
\end{equation}
The operator $\mathcal{L}_1$ is found by performing the angular integration, 
\begin{equation}
\mathcal{L}_1=\sum_{n=0}^\infty\frac{6(n+1)}{(2n+3)!}a^{2n}(\nabla^2)^n=:\frac{3}{\mathrm{i}\mathcal{D}}j_1(\mathrm{i}\mathcal{D}).
\end{equation}
Since there is no linear ambient shear flow, we find
\begin{equation}
[\boldsymbol{v}(\boldsymbol{r})-\boldsymbol{\Omega}^\infty\times\boldsymbol{r}]_\alpha=-[\mathcal{L}_1\partial_\nu G_{\alpha\beta}](\boldsymbol{r})\left(\frac{1}{2}\epsilon_{\nu\beta\lambda}\zeta^\mathrm{rr}_{\lambda\kappa}+\zeta^\mathrm{dr}_{\nu\beta\kappa}\right)[\boldsymbol{\Omega}-\boldsymbol{\Omega}^\infty]_\kappa. \label{eq:singrot}
\end{equation}
\cite{Everts:2024,Everts:2024b} showed that $[\mathcal{L}_1\nabla \mathsfbi{G}](a\hat{\boldsymbol{r}})\sim\hat{\boldsymbol{r}}$ and, therefore, all linear type boundary conditions can be satisfied. In particular,  Eq. \eqref{eq:frictionrot} can be constructed although the procedure is more complicated than the one in Sect. \ref{sec:simple}. However, with this procedure we can obtain results for a stationary sphere in general linear-ambient-flow problems, such as linear shear flow. 

For the effective pressure, we find using Eq. \eqref{eq:ind3p} that
\begin{equation}
\tilde{p}(\boldsymbol{r})-\tilde{p}^\infty=-(\nabla\boldsymbol{Q}:\boldsymbol{\zeta}^\mathrm{dr})\cdot(\boldsymbol{\Omega}-\boldsymbol{\Omega}^\infty), \label{eq:singpres}
\end{equation}
where we have used that $\nabla\times\boldsymbol{Q}(\boldsymbol{r})=\boldsymbol{0}$. Furthermore, $\tilde{p}^\infty$ (without argument) is a constant.
It is instructive to consider the $\gamma=0$ case. We then find the well-known results \citep{Kim}
\begin{equation}
\boldsymbol{v}(\boldsymbol{r})-\boldsymbol{\Omega}^\infty\times\boldsymbol{r}=-4\pi\eta_\mathrm{s}a^3\left[\nabla\times\mathsfbi{G}(\boldsymbol{r})\right]|_{\gamma=0}\cdot(\boldsymbol{\Omega}-\boldsymbol{\Omega}^\infty), \quad p(\boldsymbol{r})|_{\gamma=0}=\tilde{p}^\infty, \label{eq:strot}
\end{equation}
where we used that the Oseen tensor is symmetric. Computing the curl of the Oseen tensor, we find the rotlet velocity field $-a^3(\boldsymbol{\Omega}-\boldsymbol{\Omega}^\infty)\times\nabla\left(1/r\right)$.

To show that the same solution holds for $\gamma\neq 0$ using Eq. \eqref{eq:singrot}, we use
\begin{equation}
[\mathcal{L}_1\nabla \mathsfbi{G}]({\boldsymbol{r}})=\frac{3}{a}\int\frac{\mathrm{d}^3\boldsymbol{ k}}{(2\pi)^3}\, \mathrm{e}^{\mathrm{i}\boldsymbol{k}\cdot\boldsymbol{r}}j_1(ka)\mathrm{i}\hat{\boldsymbol{k}}\,\frac{\mathsfbi{I}-\hat{\boldsymbol{k}}\hat{\boldsymbol{k}}+\gamma (\hat{\boldsymbol{k}}\cdot\boldsymbol{\hat{\ell}})(\boldsymbol{\epsilon}\cdot\hat{\boldsymbol{k}})}{\eta_\mathrm{s}k^2[1+\gamma^2(\hat{\boldsymbol{k}}\cdot\boldsymbol{\hat{\ell}})^2]}. \label{eq:mainintegral}
\end{equation}
In contrast with the translating sphere, we do not need an explicit expression for $[\mathcal{L}_1\nabla \mathsfbi{G}]({\boldsymbol{r}})$. 
Insertion of Eq. (\ref{eq:mainintegral}) into Eq. (\ref{eq:singrot}) with friction tensors Eq. (\ref{eq:frictionrot}) results in 
\begin{gather}
\boldsymbol{v}(\boldsymbol{r})-\boldsymbol{\Omega}^\infty\times\boldsymbol{r}=-\dfrac{3a^3}{2\pi^2}(\boldsymbol{\Omega}-\boldsymbol{\Omega}^\infty)\times\nabla\underbrace{\int \mathrm{d}^3\boldsymbol{k}\, \frac{j_1(ka)}{ka}\frac{\mathrm{e}^{\mathrm{i}\boldsymbol{k}\cdot \boldsymbol{r}}}{k^2}}_{=2\pi^2/(3r)}\nonumber\\
=-a^3(\boldsymbol{\Omega}-\boldsymbol{\Omega}^\infty)\times\nabla\left(1/r\right), \label{eq:intermediate1}
\end{gather}where we used that $r>a$ for computing the integral. Eq. \eqref{eq:intermediate1} coincides with the result from Sec. \ref{sec:simple}. Using Eq. (\ref{eq:singpres}), we find for the effective pressure that
\begin{equation}
\tilde{p}(\boldsymbol{r})-\tilde{p}^\infty=-\frac{\eta_\mathrm{o}a^3}{r^3}[\boldsymbol{\hat{\ell}}-3(\boldsymbol{\hat{r}}\cdot\boldsymbol{\hat{\ell}})\boldsymbol{\hat{r}}]\cdot(\boldsymbol{\Omega}-\boldsymbol{\Omega}^\infty),
\end{equation}
which corresponds to a constant $\hat p$. The simplification that occurs for a rotating sphere does not apply to the case of a stationary sphere in an ambient linear shear flow. In this case, $[\mathcal{L}_1\nabla \mathsfbi{G}]({\boldsymbol{r}})$ needs to be computed explicitly in the fluid region, even though the results for $\boldsymbol{\zeta}^\mathrm{dr}$ and $\boldsymbol{\zeta}^\mathrm{dd}$ are known, as was (indirectly) presented by \cite{Everts:2024, Everts:2024b}. We will present the explicit expressions for the flow field for this case in our future work.

\subsection{Generalised odd-viscosity model}

The lack of an odd-viscosity contribution to the viscous dissipated power is not a generic feature of a rotating sphere. A fluid with a different type of odd viscosity than in Eq. \eqref{eq:visctensor} or a fluid with more than just one odd viscosity coefficient, generally has a $\boldsymbol{v}(\boldsymbol{r})$ that is different from the Stokes flow of a Newtonian fluid. The reason is that a general $\boldsymbol{\eta}^\mathrm{A}$ cannot be absorbed into a suitable redefinition of the pressure. Therefore, rotating (passive) spheres in general chiral active fluids dissipate more energy than in a fluid with $\boldsymbol{\eta}^\mathrm{A}=0$. The only exceptions are fluids described by the $\boldsymbol{\eta}$ given in Eq. \eqref{eq:visctensor}.

To illustrate this, we consider a minimally extended model of the viscosity tensor with two odd viscosity coefficients 

\begin{equation}
    \eta_{\alpha\beta\lambda\nu}(\boldsymbol{\hat \ell}) =\eta_\mathrm{s}\left(\delta_{\alpha\lambda}\delta_{\nu\beta}+\delta_{\alpha\nu}\delta_{\lambda\beta}-\frac{2}{3}\delta_{\alpha\beta}\delta_{\lambda\nu}\right) + \eta_\mathrm{o}^{(1)} c^{(1)}_{\alpha\beta\lambda\nu}(\boldsymbol{\hat\ell}) + \eta_\mathrm{o}^{(2)} c^{(2)}_{\alpha\beta\lambda\nu}(\boldsymbol{\hat\ell}) \label{eq:twooddvisc}
\end{equation}
where 
\begin{subeqnarray}
    c^{(1)}_{\alpha\beta\lambda\nu} (\boldsymbol{\hat\ell})&=&\hat{\ell}_\kappa(\epsilon_{\kappa\alpha\lambda}\hat\ell_\beta\hat\ell_\nu+\epsilon_{\kappa\alpha\nu}\hat\ell_\beta\hat\ell_\lambda+\epsilon_{\kappa\beta\lambda}\hat\ell_\alpha\hat\ell_\nu+\epsilon_{\kappa\beta\nu}\hat\ell_\alpha\hat\ell_\lambda),\\
    c^{(2)}_{\alpha\beta\lambda\nu}(\boldsymbol{\hat\ell})&=& \hat{\ell}_\kappa\Big[\epsilon_{\kappa\alpha\lambda}(\hat\ell_\beta\hat\ell_\nu-\delta_{\beta\nu})+\epsilon_{\kappa\alpha\nu}\hat(\ell_\beta\hat\ell_\lambda-\delta_{\beta\lambda})\nonumber\\
    &&+\epsilon_{\kappa\beta\lambda}(\hat\ell_\alpha\hat\ell_\nu-\delta_{\alpha\nu})
+\epsilon_{\kappa\beta\nu}(\hat\ell_\alpha\hat\ell_\lambda-\delta_{\alpha\lambda})\Big].
\end{subeqnarray}
Furthermore, $\eta_\mathrm{o}^{(1)}$ and $\eta_\mathrm{o}^{(2)}$ are odd-viscosity coefficients.
Inserting Eq. \eqref{eq:twooddvisc} into the linear momentum balance Eq. \eqref{eq:creeping} for $\boldsymbol{f=0}$ gives
\begin{gather}
    \eta_s\nabla^2\boldsymbol{v}(\boldsymbol{r}) -\nabla\hat p(\boldsymbol{r}) -\left(\eta_\mathrm{o}^{(1)}+2\eta_\mathrm{o}^{(2)}\right)\nabla^2[\boldsymbol{v}(\boldsymbol{r})\times \boldsymbol{\hat\ell}] 
    \nonumber\\
    +2\left(\eta_\mathrm{o}^{(1)}+\eta_\mathrm{o}^{(2)}\right)(\boldsymbol{\hat\ell}\cdot\nabla)^2[\boldsymbol{v}(\boldsymbol{r})\times\boldsymbol{\hat\ell}] = \boldsymbol{0}, \label{eq:twoodd}
\end{gather}
where the effective pressure takes the form $\hat p(\boldsymbol{r})=p(\boldsymbol{r})-\eta_\mathrm{o}^{(1)}\boldsymbol{\hat\ell}\cdot[\nabla\times \boldsymbol{v}(\boldsymbol{r})]$. This extended model reduces to the one odd-viscosity model from Eq. \eqref{eq:visctensor}, when $\eta_\mathrm{o}^{(2)}=-\eta_\mathrm{o}^{(1)}=\eta_\mathrm{o}$. The additional term $(\boldsymbol{\hat\ell}\cdot\nabla)^2[\boldsymbol{v}(\boldsymbol{r})\times\boldsymbol{\hat\ell}]$ can manifestly not be absorbed into a modified pressure. We conclude that for $\eta_\mathrm{o}^{(2)}\neq-\eta_\mathrm{o}^{(1)}$, the class of solutions to the ordinary Stokes equations listed in Eq. \eqref{eq:multipole} are not solutions to Eq. \eqref{eq:twoodd}. In particular, this observation implies that the rotlet is not a solution. It follows from our generalised Helmholtz theorem, that generally there are contributions of $\eta_\mathrm{o}^{(1)}$ and $\eta_\mathrm{o}^{(2)}$ to the viscous dissipation rate of a rotating sphere and the value of the dissipated power is higher than for a Newtonian fluid.

\section{Conclusions}
In summary, we have shown that the creeping flow equations for a general viscosity tensor admit a unique solution for the fluid flow. Furthermore, we have proven a general Helmholtz theorem showing that odd fluids generally dissipate more energy than equivalent fluids without odd viscosity, unless both systems share the same fluid velocity field. An example of higher viscous dissipation due to odd effects is a translating sphere in an odd fluid with viscosity tensor Eq. \eqref{eq:visctensor}, whereas an example of equal dissipation is the rotating sphere. For both systems, we have derived exact singularity representations of the velocity and pressure fields, for which explicit closed-form expressions were obtained. As in previous works \citep{Khain:2022, Everts:2024}, we retrieve the axial flow fields for a translating sphere (not just restricted to small $\eta_\mathrm{o}$), and have discussed how the flow field is altered when the direction of spin angular momentum is not aligned with the translation direction of the particle. Compared to a Newtonian fluid, the rotating sphere in an odd fluid has only a modified pressure, but the same type of rotlet flow field. Finally, we derived an exact expression for the stress tensor of the fundamental solution, which can be used to numerically compute mobility (or equivalently, friction) tensors of arbitrarily shaped particles subjected to general boundary conditions. Our results are important for the development of odd microhydrodynamics, microswimmers suspended in odd fluids, and odd Brownian motion. In future work, we will focus on the flow fields produced by a particle in an ambient linear straining flow, for which the dipolar-dipolar sector takes an important role.

\begin{bmhead}[Acknowledgements.]
We thank Andrej Vilfan and Simon Čopar for insightful discussions.
\end{bmhead}

\begin{bmhead}[Funding.]
We acknowledge funding from the National Science Centre, Poland, within the OPUS LAP grant no. 2024/55/I/ST3/00998.
\end{bmhead}

\begin{bmhead}[Declaration of interests.]
The authors report no conflict of interest.
\end{bmhead}
\newpage

\begin{appen}
\section{}\label{app}
In this Appendix we list the tensor components for $\boldsymbol{\Sigma}(\boldsymbol{r})$ defined in Eq. \eqref{eq:fundstress}, which can be determined from Eq. \eqref{eq:greenodd}. 
The strain rate tensor is (using the short-hand notation $\Lambda=\Lambda(s)$),
\begin{align}
&8\pi\eta_\mathrm{s}\Theta_{\alpha\beta\kappa}(\boldsymbol{r})=4\pi\eta_\mathrm{s}\left[\partial_\alpha G_{\beta\kappa}(\boldsymbol{r})+\partial_\beta G_{\alpha\kappa}(\boldsymbol{r})\right] \nonumber\\
&=\frac{\Lambda}{r^2(1+\Lambda)}\Big[-(\hat{r}_\alpha\delta_{\beta\kappa}+\hat{r}_\beta\delta_{\alpha\kappa})+\Lambda(2\delta_{\alpha\beta}\hat{r}_\kappa+\delta_{\alpha\kappa}\hat{r}_\beta+\delta_{\beta\kappa}\hat{r}_\alpha-6\hat{r}_\alpha\hat{r}_\beta\hat{r}_\kappa)\nonumber\\
&+\Lambda(\Lambda-1)(\Lambda+2)(b_\alpha\hat{r}_\beta\hat{r}_\kappa+\hat{r}_\alpha b_\beta\hat{r}_\kappa)+\left[2-2\Lambda+\Lambda(\Lambda-1)(\Lambda+2)\right](b_\alpha \hat{\phi}_\beta \hat{\phi}_\kappa+\hat{\phi}_\alpha b_\beta \hat{\phi}_\kappa) \nonumber\\
&+(1-\Lambda)\left(2\hat{\phi}_\alpha \hat{\phi}_\beta b_\kappa-b_\alpha\delta_{\beta\kappa}-b_\beta\delta_{\alpha\kappa}+ 2\hat{\phi}_\alpha\hat{r}_\beta \hat{\phi}_\kappa+2 \hat{\phi}_\alpha \hat{\phi}_\beta\hat{r}_\kappa+2\hat{r}_\alpha \hat{\phi}_\beta \hat{\phi}_\kappa\right) \nonumber
\\
&+s\Lambda\left(\Lambda^2+\Lambda-1\right)(b_\alpha\hat{r}_\beta \hat{\phi}_\kappa+\hat{r}_\alpha b_\beta \hat{\phi}_\kappa-b_\alpha \hat{\phi}_\beta\hat{r}_\kappa- \hat{\phi}_\alpha b_\beta\hat{r}_\kappa)\nonumber\\
&+s\Lambda(2\delta_{\alpha\beta} \hat{\phi}_\kappa-\delta_{\alpha\kappa} \hat{\phi}_\beta-\delta_{\beta\kappa} \hat{\phi}_\alpha+2\hat{r}_\alpha \hat{\phi}_\beta\hat{r}_\kappa+2\hat{\phi}_\alpha\hat{r}_\beta \hat{r}_\kappa-4\hat{r}_\alpha\hat{r}_\beta \hat{\phi}_\kappa+ \hat{\phi}_\alpha b_\beta\hat{r}_\kappa+ b_\alpha\hat{\phi}_\beta \hat{r}_\kappa\nonumber\\
&- \hat{\phi}_\alpha\hat{r}_\beta b_\kappa- \hat{r}_\alpha\hat{\phi}_\beta b_\kappa)
\Big],
\end{align}
with $\boldsymbol{b}=(\boldsymbol{\hat{r}}\cdot\boldsymbol{\hat{\ell}})/{|\boldsymbol{\hat{r}}\times\boldsymbol{\hat{\ell}}|}\boldsymbol{\hat{\theta}}$. Furthermore, we have
\begin{align}
4\pi\eta_\mathrm{s}[\nabla\times\mathsfbi{G}(\boldsymbol{r})]=&\frac{\Lambda}{r^2}\Bigg((\boldsymbol{\hat{\theta}}\boldsymbol{\hat{\phi}}-\boldsymbol{\hat{\phi}}\boldsymbol{\hat{\theta}})-\frac{\boldsymbol{\hat{r}}\cdot\boldsymbol{\hat{\ell}}}{|\boldsymbol{\hat{r}}\times\boldsymbol{\hat{\ell}}|}s^2\Lambda^2(\boldsymbol{\hat{r}}\boldsymbol{\hat{\phi}}-\boldsymbol{\hat{\phi}}\boldsymbol{\hat{r}})\nonumber\\
&-\frac{s\Lambda}{1+\Lambda}\left\{\boldsymbol{\hat{\theta}}\boldsymbol{\hat{r}}+\boldsymbol{\hat{r}}\boldsymbol{\hat{\theta}}+\frac{\boldsymbol{\hat{r}}\cdot\boldsymbol{\hat{\ell}}}{|\boldsymbol{\hat{r}}\times\boldsymbol{\hat{\ell}}|}\left[\boldsymbol{\hat{\theta}}\boldsymbol{\hat{\theta}}-\boldsymbol{\hat{\phi}}\boldsymbol{\hat{\phi}}+\Lambda(\Lambda+1)(\mathsfbi{I}-\boldsymbol{\hat{\theta}}\boldsymbol{\hat{\theta}})\right]\right\}\Bigg),
\end{align}
which leads to the explicit form

\begin{equation}
\boldsymbol{P}(\boldsymbol{r})=\frac{1}{4\pi r^2}\left[2\Lambda^3(1+\gamma^2)(\boldsymbol{\hat{r}}+s\boldsymbol{\hat{\phi}})-\boldsymbol{\hat{r}}\right].
\end{equation}

\section{}\label{appA}
In this Appendix we give details on evaluating Eq. \eqref{eq:Fourierrep}. We write
\begin{equation}\mathcal{L}_0\mathsfbi{G}(\boldsymbol{r})=\mathrm{Tr}[\mathsfbi{B}(\boldsymbol{r})]\mathsfbi{I}-\mathsfbi{B}(\boldsymbol{r})+\gamma\boldsymbol{\epsilon}\cdot\mathsfbi{B}(\boldsymbol{r})\cdot\boldsymbol{\hat{\ell}}\label{eq:lnotG}\end{equation} where the tensor $\mathsfbi{B}(\boldsymbol{r})$ is given by
\begin{equation}
\mathsfbi{B}(\boldsymbol{r})=\frac{1}{\eta_\mathrm{s}}\int \frac{\mathrm{d}^3\boldsymbol{k}}{(2\pi)^3}\, \mathrm{e}^{\mathrm{i}\boldsymbol{k}\cdot\boldsymbol{r}}j_0(ka)\frac{\hat{\boldsymbol{k}}\hat{\boldsymbol{k}}}{k^2[1+\gamma^2(\hat{\boldsymbol{k}}\cdot\boldsymbol{\hat{\ell}})^2]}. \label{eq:Btensor}
\end{equation}
We define a cylindrical coordinate system by: $k_x=k_\perp\cos k_\phi$, $k_y=k_\perp\sin k_\phi, k_z=\boldsymbol{k}\cdot\boldsymbol{\hat{\ell}}$ and $x=\rho\cos\phi$, $y=\rho\sin\phi$, $z=z$. Using these coordinates, Eq. (\ref{eq:Btensor}) reduces to
\begin{equation}
\mathsfbi{B}(\boldsymbol{r})=\frac{1}{(2\pi)^3\eta_\mathrm{s}}\int_0^\infty \mathrm{d}k_\perp\, k_\perp \int_0^{2\pi}\mathrm{d}k_\phi\, \mathrm{e}^{\mathrm{i}k_\perp\rho\cos(k_\phi-\phi)}\int_{-\infty}^\infty \mathrm{d}k_z\, \frac{j_0\left(a\sqrt{k_\perp^2+k_z^2}\right)k_\alpha k_\beta \mathrm{e}^{\mathrm{i}k_zz}}{[k_\perp^2+(1+\gamma^2)k_z^2](k_\perp^2+k_z^2)}.
\end{equation}
The integrals over $k_z$ and $k_\phi$ can be  evaluated explicitly by contour integration and using results from \cite{Gradsteyn}, respectively. See \cite{Everts:2024} for further details. The result is
\begin{eqnarray}
\mathsfbi{B}(\boldsymbol{r})=\int_0^\infty \frac{\mathrm{d}k_\perp}{{4\pi\eta_\mathrm{s}\gamma^2}}\Bigg\{L_{0}(k_\perp,z)\frac{J_1(k_\perp\rho)}{k_\perp\rho}\boldsymbol{\hat{\phi}}\boldsymbol{\hat{\phi}}+L_0(k_\perp,z)\left[J_0(k_\perp \rho)-\frac{J_1(k_\perp\rho)}{k_\perp\rho}\right]\boldsymbol{\hat{\rho}}\boldsymbol{\hat{\rho}}\nonumber \\-L_1(k_\perp,z)J_1(k_\perp\rho)(\boldsymbol{\hat{\rho}}\boldsymbol{\hat{\ell}}+\boldsymbol{\hat{\ell}}\boldsymbol{\hat{\rho}})
-L_2(k_\perp,z)J_0(k_\perp \rho)\boldsymbol{\hat{\ell}}\boldsymbol{\hat{\ell}}\Bigg\},  \label{eq:lastone}
\end{eqnarray}
with
\begin{equation}
L_{m}(k_\perp,z)=[\mathrm{sgn}(z)]^m\left[\mathrm{e}^{-k_\perp\cos\psi |z|}j_0(k_\perp a\sin\psi)(\cos\psi)^{m-1}- \mathrm{e}^{-k_\perp|z|}\right], \quad m=0,1,2.\label{eq:M}
\end{equation}
Each term in Eq. (\ref{eq:lastone}) can be explicitly computed using the following integrals \citep{Gradsteyn}:
\begin{align}
   \int_0^\infty \mathrm{d}k_\perp\, \mathrm{e}^{-k_\perp|z|}\frac{J_1(k_\perp\rho)}{ k_\perp\rho}=&\dfrac{\sqrt{\rho^2+z^2}-|z|}{\rho^2}, \label{eq:bessela}\\
   \int_0^\infty \mathrm{d}k_\perp\, \mathrm{e}^{-k_\perp|z|}J_m(k_\perp\rho)=&\frac{\left(\sqrt{\rho^2+z^2}-|z|\right)^m}{\rho^m\sqrt{\rho^2+z^2}}, \quad m=0,1,...,\\
    \int_0^\infty \mathrm{d}k_\perp\, \mathrm{e}^{-k_\perp|z| \cos\psi}J_0(k_\perp\rho)j_0(k_\perp a\sin\psi)=&\frac{1}{a\sin\psi}\mathrm{arcsin}\left[\frac{1}{\mathcal{R}_+(\boldsymbol{r};\gamma)}\right],\\
    \int_0^\infty \mathrm{d}k_\perp\, \mathrm{e}^{-k_\perp|z|\cos\psi}J_1(k_\perp\rho)j_0(k_\perp a\sin\psi)=&
    \frac{1}{\rho}\left[1-\sqrt{1-\mathcal{R}_-(\boldsymbol{r};\gamma)^2}\right],\\
    \int_0^\infty \mathrm{d}k_\perp\, \mathrm{e}^{-k_\perp|z|\cos\psi}\frac{J_1(k_\perp\rho)}{ k_\perp\rho}j_0(k_\perp a\sin\psi)=&\frac{1}{2a\sin\psi}\mathrm{arcsin}\left[\frac{1}{\mathcal{R}_+(\boldsymbol{r};\gamma)}\right] \label{eq:besselz} \\
   &+\frac{a\sin\psi}{2\rho^2}\sqrt{\mathcal{R}_+(\boldsymbol{r};\gamma)^2-1}\left[1-\sqrt{1-\mathcal{R}_-(\boldsymbol{r};\gamma)^2}\right]^2. \nonumber 
\end{align}
where the quantities $\mathcal{R}_\pm$ are defined in Eq. \eqref{eq:rpm} with the identification $\rho=r|\boldsymbol{\hat{r}}\times\boldsymbol{\hat{\ell}}|$ and $z=r(\boldsymbol{\hat{r}}\cdot\boldsymbol{\hat{\ell}})$. Furthermore, we used that $\rho>0$ and $\cos[\psi(\gamma)]>0$. 
Insertion of Eqs. \eqref{eq:bessela}-\eqref{eq:besselz} into Eqs. \eqref{eq:lastone} and \eqref{eq:M} gives $\mathsfbi{B}(\boldsymbol{r})$. From $\mathsfbi{B}(\boldsymbol{r})$ and Eq. (\ref{eq:lnotG}), it follows that 
\begin{align}
\mathcal{L}_0\mathsfbi{G}(\boldsymbol{r})=&\frac{1}{4\pi\eta_\mathrm{s}a\gamma^2}\Bigg\{\left[\frac{1+\gamma^2}{\gamma}\mathcal{M}(\boldsymbol{r};\gamma)-\frac{a}{r}\right]\boldsymbol{\hat{\ell}}\boldsymbol{\hat{\ell}}-\left[\frac{1}{2}\mathcal{O}(\boldsymbol{r};\gamma)+\frac{1-\gamma^2}{2\gamma}\mathcal{M}(\boldsymbol{r};\gamma)-\frac{a}{2 r}\right](\mathsfbi{I}-\boldsymbol{\hat{\ell}}\boldsymbol{\hat{\ell}}) \nonumber \\
&+\mathcal{O}(\boldsymbol{r};\gamma)\boldsymbol{\hat{\rho}}\boldsymbol{\hat{\rho}}+\mathcal{N}(\boldsymbol{r};\gamma)\left[\gamma(\boldsymbol{\epsilon}\cdot\boldsymbol{\hat{\rho}})-(\boldsymbol{\hat{\rho}}\boldsymbol{\hat{\ell}}+\boldsymbol{\hat{\ell}}\boldsymbol{\hat{\rho}})\right]-\left[\mathcal{M}(\boldsymbol{r};\gamma)-\frac{a\gamma}{r}\right](\boldsymbol{\epsilon}\cdot\boldsymbol{\hat{\ell}})\Bigg\},
\label{eq:Lauraresult}
\end{align}
with $\mathcal{M}(\boldsymbol{r};\gamma)$, $\mathcal{N}(\boldsymbol{r};\gamma)$, and $\mathcal{O}(\boldsymbol{r};\gamma)$ defined in Eq. \eqref{eq:something}. 

It is instructive to evaluate Eq. \eqref{eq:Lauraresult} on the surface of a sphere with radius $a$. It is straightforward to check that
\begin{equation}
\mathcal{R}_+(a\boldsymbol{\hat{r}};\gamma)=\frac{1}{\sin\psi},\quad \mathcal{R}_-(a\boldsymbol{\hat{r}};\gamma)=|\boldsymbol{\hat{r}}\times\boldsymbol{\hat{\ell}}|,
\end{equation}
and, therefore, $\mathcal{M}(a\boldsymbol{\hat{r}};\gamma)=\mathrm{arctan}(\gamma)$ and $\mathcal{N}(a\boldsymbol{\hat{r}};\gamma)=\mathcal{O}(a\boldsymbol{\hat{r}};\gamma)=0$, where we used that $\gamma>0$. Direct substitution of these results in Eq. \eqref{eq:Lauraresult} gives
\begin{equation}
\mathcal{L}_0\mathsfbi{G}(a\boldsymbol{\boldsymbol{\hat{r}}})=\frac{1}{24\pi\eta_\mathrm{s}a}\left[R(\gamma)(\mathsfbi{I}-\boldsymbol{\hat{\ell}}\boldsymbol{\hat{\ell}})+T(\gamma)\boldsymbol{\hat{\ell}}\boldsymbol{\hat{\ell}}-S(\gamma)(\boldsymbol{\epsilon}\cdot\boldsymbol{\hat{\ell}})\right],
\end{equation}
which equals the translational-translational mobility tensor $\boldsymbol{\mu}^\mathrm{tt}=[\boldsymbol{\zeta}^\mathrm{tt}]^{-1}$, as it should be.
\end{appen}

\bibliographystyle{jfm}
\bibliography{jfm}

@article{Hosaka:2024,
  title = {Chirotactic response of microswimmers in fluids with odd viscosity},
  author = {Hosaka, Y. and Chatzittofi, M. and Golestanian, R. and Vilfan, A.},
  journal = {Phys. Rev. Res.},
  volume = {6},
  issue = {3},
  pages = {L032044},
  numpages = {8},
  year = {2024},
  month = {Aug},
  publisher = {American Physical Society},
  doi = {10.1103/PhysRevResearch.6.L032044},
  url = {https://link.aps.org/doi/10.1103/PhysRevResearch.6.L032044}
}

@article{Everts:2024,
  title = {Dissipative Effects in Odd Viscous {S}tokes Flow around a Single Sphere},
  author = {Everts, J. C. and Cichocki, B.},
  journal = {Phys. Rev. Lett.},
  volume = {132},
  issue = {21},
  pages = {218303},
  numpages = {7},
  year = {2024},
  month = {May},
  publisher = {American Physical Society},
  doi = {10.1103/PhysRevLett.132.218303},
  url = {https://link.aps.org/doi/10.1103/PhysRevLett.132.218303}
}

@article{Lapa:2014,
  title = {Swimming at low {R}eynolds number in fluids with odd, or {H}all, viscosity},
  author = {Lapa, M. F. and Hughes, T. L.},
  journal = {Phys. Rev. E},
  volume = {89},
  issue = {4},
  pages = {043019},
  numpages = {13},
  year = {2014},
  month = {Apr},
  publisher = {American Physical Society},
  doi = {10.1103/PhysRevE.89.043019},
  url = {https://link.aps.org/doi/10.1103/PhysRevE.89.043019}
}

@article{Mazur:1974,
title = {A generalization of {F}axén's theorem to nonsteady motion of a sphere through an incompressible fluid in arbitrary flow},
journal = {Physica},
volume = {76},
number = {2},
pages = {235-246},
year = {1974},
issn = {0031-8914},
doi = {https://doi.org/10.1016/0031-8914(74)90197-9},
author = {P. Mazur and D. Bedeaux}
}

@article{Khain:2022, title={Stokes flows in three-dimensional fluids with odd and parity-violating viscosities}, volume={934}, DOI={10.1017/jfm.2021.1079}, journal={J. Fluid Mech.}, publisher={Cambridge University Press}, author={Khain, T. and Scheibner, C. and Fruchart, M. and Vitelli, V.}, year={2022}, pages={A23}}

@article{Cruz:2023,
  title = {Stokesian dynamics with odd viscosity},
  author = {Yuan, H. and Olvera de la Cruz, M.},
  journal = {Phys. Rev. Fluids},
  volume = {8},
  issue = {5},
  pages = {054101},
  numpages = {29},
  year = {2023},
  month = {May},
  publisher = {American Physical Society},
  doi = {10.1103/PhysRevFluids.8.054101},
  url = {https://link.aps.org/doi/10.1103/PhysRevFluids.8.054101}
}

@book{Gradsteyn,
  title={Table of integrals, series, and products},
  author={Gradshteyn, I. S. and Ryzhik, I. M.},
  year={2014},
  publisher={Academic press}
}

@article{Souslov:2019,
  title = {Topological Waves in Fluids with Odd Viscosity},
  author = {Souslov, A. and Dasbiswas, K. and Fruchart, M. and Vaikuntanathan, S. and Vitelli, V.},
  journal = {Phys. Rev. Lett.},
  volume = {122},
  issue = {12},
  pages = {128001},
  numpages = {6},
  year = {2019},
  month = {Mar},
  publisher = {American Physical Society},
  doi = {10.1103/PhysRevLett.122.128001},
  url = {https://link.aps.org/doi/10.1103/PhysRevLett.122.128001}
}

@article{Chen:2024,
  title = {Odd Viscosity Suppresses Intermittency in Direct Turbulent Cascades},
  author = {Chen, S. and de Wit, X. M. and Fruchart, M. and Toschi, F. and Vitelli, V.},
  journal = {Phys. Rev. Lett.},
  volume = {133},
  issue = {14},
  pages = {144002},
  numpages = {6},
  year = {2024},
  month = {Sep},
  publisher = {American Physical Society},
  doi = {10.1103/PhysRevLett.133.144002},
  url = {https://link.aps.org/doi/10.1103/PhysRevLett.133.144002}
}

@article{Wit:2024,
  title={Pattern formation by turbulent cascades},
  author={de Wit, X. M. and Fruchart, M. and Khain, T. and Toschi, F. and Vitelli, V.},
  journal={Nature},
  volume={627},
  number={8004},
  pages={515--521},
  year={2024},
  publisher={Nature Publishing Group UK London},
  doi={10.1038/s41586-024-07074-z}
}

@article{Lou:2022,
author = {X. Lou  and Q. Yang  and Y. Ding  and P. Liu  and K. Chen  and X. Zhou  and F. Ye  and R. Podgornik  and M. Yang },
title = {Odd viscosity-induced {H}all-like transport of an active chiral fluid},
journal = {Proc. Natl. Acad. Sci. U.S.A.},
volume = {119},
number = {42},
pages = {e2201279119},
year = {2022},
doi = {10.1073/pnas.2201279119}}

@article{
Geim:2019,
author = {A. I. Berdyugin  and S. G. Xu  and F. M. D. Pellegrino  and R. Krishna Kumar  and A. Principi  and I. Torre  and M. Ben Shalom  and T. Taniguchi  and K. Watanabe  and I. V. Grigorieva  and M. Polini  and A. K. Geim  and D. A. Bandurin },
title = {Measuring {H}all viscosity of graphene’s electron fluid},
journal = {Science},
volume = {364},
number = {6436},
pages = {162-165},
year = {2019},
doi = {10.1126/science.aau0685}}

@article{Soni:2019,
  title={The odd free surface flows of a colloidal chiral fluid},
  author={Soni, V. and Bililign, E. S. and Magkiriadou, S. and Sacanna, S. and Bartolo, D. and Shelley, M. J. and Irvine, W. T. M.},
  journal={Nat. Phys.},
  volume={15},
  number={11},
  pages={1188--1194},
  year={2019},
  publisher={Nature Publishing Group UK London},
 doi={10.1038/s41567-019-0603-8}
}

@article{Beenakker:1971,
  title = {Influence of a Magnetic Field on the Transport Coefficients of Oxygen Gas: Anomalies Associated with the $\ensuremath{\sigma}=0$ Multiplets},
  author = {Beenakker, J. J. M. and Coope, J. A. R. and Snider, R. F.},
  journal = {Phys. Rev. A},
  volume = {4},
  issue = {2},
  pages = {788--796},
  numpages = {0},
  year = {1971},
  month = {Aug},
  publisher = {American Physical Society},
  doi = {10.1103/PhysRevA.4.788},
  url = {https://link.aps.org/doi/10.1103/PhysRevA.4.788}
}

@article{Callen:1992,
    author = {Chang, Z. and Callen, J. D.},
    title = {Generalized gyroviscous force and its effect on the momentum balance equation},
    journal = {Phys. Fluids B},
    volume = {4},
    number = {7},
    pages = {1766-1771},
    year = {1992},
    month = {07},
    issn = {0899-8221},
    doi = {10.1063/1.860032}
}

@book{Groot,
  title={Non-equilibrium thermodynamics},
  author={De Groot, S. R. and Mazur, P.},
  year={1962},
  publisher={Amsterdam, Holland: North-Holland Publishing Company.}
}

@article{Onsager:1931a,
  title = {Reciprocal Relations in Irreversible Processes. {I}.},
  author = {Onsager, L.},
  journal = {Phys. Rev.},
  volume = {37},
  issue = {4},
  pages = {405--426},
  numpages = {0},
  year = {1931},
  month = {Feb},
  publisher = {American Physical Society},
  doi = {10.1103/PhysRev.37.405},
  url = {https://link.aps.org/doi/10.1103/PhysRev.37.405}
}

@article{Onsager:1931b,
  title = {Reciprocal Relations in Irreversible Processes. {II}.},
  author = {Onsager, L.},
  journal = {Phys. Rev.},
  volume = {38},
  issue = {12},
  pages = {2265--2279},
  numpages = {0},
  year = {1931},
  month = {Dec},
  publisher = {American Physical Society},
  doi = {10.1103/PhysRev.38.2265},
  url = {https://link.aps.org/doi/10.1103/PhysRev.38.2265}
}

@article{Casimir:1945,
  title = {On {O}nsager's Principle of Microscopic Reversibility},
  author = {Casimir, H. B. G.},
  journal = {Rev. Mod. Phys.},
  volume = {17},
  issue = {2-3},
  pages = {343--350},
  numpages = {0},
  year = {1945},
  month = {Apr},
  publisher = {American Physical Society},
  doi = {10.1103/RevModPhys.17.343},
  url = {https://link.aps.org/doi/10.1103/RevModPhys.17.343}
}

@article{Fruchart:2023,
author = {Fruchart, M. and Scheibner, C. and Vitelli, V.},
title = {Odd Viscosity and Odd Elasticity},
journal = {Annu. Rev. Condens. Matter Phys.},
volume = {14},
number = {1},
pages = {471-510},
year = {2023},
doi = {10.1146/annurev-conmatphys-040821-125506}
}

@article{Sumino:2012,
  title={Large-scale vortex lattice emerging from collectively moving microtubules},
  author={Sumino, Y. and Nagai, K. H. and Shitaka, Y. and Tanaka, D. and Yoshikawa, K. and Chat{\'e}, H. and Oiwa, K.},
  journal={Nature},
  volume={483},
  number={7390},
  pages={448--452},
  year={2012},
  publisher={Nature Publishing Group UK London},
doi={10.1038/nature10874}
}

@article{
Cartwright:2004,
author = {J. H. E. Cartwright  and O. Piro  and I. Tuval },
title = {Fluid-dynamical basis of the embryonic development of left-right asymmetry in vertebrates},
journal = {Proc. Natl. Acad. Sci. U.S.A.},
volume = {101},
number = {19},
pages = {7234-7239},
year = {2004},
doi = {10.1073/pnas.0402001101}}

@article{
Riedel:2005,
author = {I. H. Riedel  and K. Kruse  and J. Howard },
title = {A Self-Organized Vortex Array of Hydrodynamically Entrained Sperm Cells},
journal = {Science},
volume = {309},
number = {5732},
pages = {300-303},
year = {2005},
doi = {10.1126/science.1110329}}

@article{
Scholz:2021,
author = {C. Scholz  and A. Ldov  and T. Pöschel  and M. Engel  and H. Löwen },
title = {Surfactants and rotelles in active chiral fluids},
journal = {Sci. Adv.},
volume = {7},
number = {16},
pages = {eabf8998},
year = {2021},
doi = {10.1126/sciadv.abf8998}}

@article{Avron:1988,
  title={Odd viscosity},
  author={Avron, J.E.},
  journal={J. Stat. Phys},
  volume={92},
  pages={543--557},
  year={1998},
  doi={10.1023/A:1023084404080},
  publisher={Springer}
}

@article{Lier:2024, title={Odd viscous flow past a sphere at low but non-zero {R}eynolds numbers}, volume={998}, DOI={10.1017/jfm.2024.915}, journal={J. Fluid Mech.}, author={Lier, R.}, year={2024}, pages={A40}}

@article{Ganeshan:2017,
  title = {Odd viscosity in two-dimensional incompressible fluids},
  author = {Ganeshan, Sriram and Abanov, Alexander G.},
  journal = {Phys. Rev. Fluids},
  volume = {2},
  issue = {9},
  pages = {094101},
  numpages = {13},
  year = {2017},
  month = {Sep},
  publisher = {American Physical Society},
  doi = {10.1103/PhysRevFluids.2.094101},
  url = {https://link.aps.org/doi/10.1103/PhysRevFluids.2.094101}
}

@article{Lier:2023,
  title = {Lift force in odd compressible fluids},
  author = {Lier, R. and Duclut, C. and Bo, S. and Armas, J. and J\"ulicher, F. and Sur\'owka, P.},
  journal = {Phys. Rev. E},
  volume = {108},
  issue = {2},
  pages = {L023101},
  numpages = {7},
  year = {2023},
  month = {Aug},
  publisher = {American Physical Society},
  doi = {10.1103/PhysRevE.108.L023101},
  url = {https://link.aps.org/doi/10.1103/PhysRevE.108.L023101}
}

@article{Hosaka:2025,
    author = {Daddi-Moussa-Ider, A. and Vilfan, A. and Hosaka, Y.},
    title = {Analytical solution for the hydrodynamic resistance of a disk in a compressible fluid layer with odd viscosity on a rigid substrate},
    journal = {J. Chem. Phys.},
    volume = {162},
    number = {6},
    pages = {064103},
    year = {2025},
    month = {02},
    issn = {0021-9606},
    doi = {10.1063/5.0249623}
}

@article{Khain:2024, title={Trading particle shape with fluid symmetry: on the mobility matrix in 3-D chiral fluids}, volume={992}, DOI={10.1017/jfm.2024.535}, journal={J. Fluid Mech.}, author={Khain, T. and Fruchart, M. and Scheibner, C. and Witten, T. A. and Vitelli, V.}, year={2024}, pages={A5}}

@book{Kim,
  title={Microhydrodynamics: principles and selected applications},
  author={Kim, S. and Karrila, S. J.},
  year={2013},
  publisher={Butterworth-Heinemann}
}

@book{pozrikidis1992,
        Address = {},
        Author = {Pozrikidis, C.},
        Publisher = {Cambridge University Press},
        Doi = {https://doi.org/10.1017/CBO9780511624124},
        Title = {Boundary integral and singularity methods for linearized viscous flow},
        Year = {1992}}

@article{Brenner:1966,
title = {The {S}tokes resistance of an arbitrary particle—Part {V}.: Symbolic operator representation of intrinsic resistance},
journal = {Chem. Eng. Sci.},
volume = {21},
number = {1},
pages = {97-109},
year = {1966},
issn = {0009-2509},
doi = {10.1016/0009-2509(66)80010-6},
author = {H. Brenner}}

@article{Brenner:1964,
title = {The {S}tokes resistance of an arbitrary particle—{IV} Arbitrary fields of flow},
journal = {Chem. Eng. Sci.},
volume = {19},
number = {10},
pages = {703-727},
year = {1964},
issn = {0009-2509},
doi = {https://doi.org/10.1016/0009-2509(64)85084-3},
url = {https://www.sciencedirect.com/science/article/pii/0009250964850843},
author = {H. Brenner}
}

@article{Brenner:1964b,
title = {The {S}tokes resistance of a slightly deformed sphere},
journal = {Chem. Eng. Sci.},
volume = {19},
number = {8},
pages = {519-539},
year = {1964},
issn = {0009-2509},
doi = {https://doi.org/10.1016/0009-2509(64)85045-4},
url = {https://www.sciencedirect.com/science/article/pii/0009250964850454},
author = {H. Brenner}
}

@book{HappelBrenner,
  title={Low {Reynolds} number hydrodynamics: with special applications to particulate media},
  author={Happel, John and Brenner, Howard},
  year={1983},
  publisher={Springer Science \& Business Media}
}

@article{Helmholtz,
  title={Zur Theorie der station{\"a}ren Str{\"o}me in reibenden Fl{\"u}ssigkeiten},
  author={Helmholtz, H.},
  journal={Wiss. Abh},
  volume={1},
  pages={223--230},
  year={1868}
}

@article{Vilfan:2023,
  title = {Lorentz Reciprocal Theorem in Fluids with Odd Viscosity},
  author = {Hosaka, Y. and Golestanian, R. and Vilfan, A.},
  journal = {Phys. Rev. Lett.},
  volume = {131},
  issue = {17},
  pages = {178303},
  numpages = {7},
  year = {2023},
  month = {Oct},
  publisher = {American Physical Society},
  doi = {10.1103/PhysRevLett.131.178303},
  url = {https://link.aps.org/doi/10.1103/PhysRevLett.131.178303}
}

@article{Banerjee:2017,
  title={Odd viscosity in chiral active fluids},
  author={Banerjee, D. and Souslov, A. and Abanov, A. G. and Vitelli, V.},
  journal={Nat. Commun.},
  volume={8},
  number={1},
  pages={1573},
  doi={10.1038/s41467-017-01378-7},
  year={2017},
  publisher={Nature Publishing Group UK London}
}

@article{Markovich:2021,
  title = {Odd Viscosity in Active Matter: Microscopic Origin and {3D} Effects},
  author = {Markovich, T. and Lubensky, T. C.},
  journal = {Phys. Rev. Lett.},
  volume = {127},
  issue = {4},
  pages = {048001},
  numpages = {6},
  year = {2021},
  month = {Jul},
  publisher = {American Physical Society},
  doi = {10.1103/PhysRevLett.127.048001},
  url = {https://link.aps.org/doi/10.1103/PhysRevLett.127.048001}
}

@article{Groot:1964,
  title = {Extension of {O}nsager's Theory of Reciprocal Relations. I},
  author = {de Groot, S. R. and Mazur, P.},
  journal = {Phys. Rev.},
  volume = {94},
  issue = {2},
  pages = {218--224},
  numpages = {0},
  year = {1954},
  month = {Apr},
  publisher = {American Physical Society},
  doi = {10.1103/PhysRev.94.218},
  url = {https://link.aps.org/doi/10.1103/PhysRev.94.218}
}

@article{Everts:2024b,
  title = {{E}rratum: Dissipative Effects in Odd Viscous {S}tokes Flow around a Single Sphere [{P}hys. {R}ev. {L}ett. 132, 218303 (2024)]},
  author = {Everts, J. C. and Cichocki, B.},
  journal = {Phys. Rev. Lett.},
  volume = {133},
  issue = {3},
  pages = {039902},
  numpages = {1},
  year = {2024},
  month = {Jul},
  publisher = {American Physical Society},
  doi = {10.1103/PhysRevLett.133.039902},
  url = {https://link.aps.org/doi/10.1103/PhysRevLett.133.039902}
}

@article{Cichocki:1988,
author = {Cichocki, B. and Felderhof, U. and Schmitz, R.},
year = {1988},
month = {01},
pages = {383-403},
title = {Hydrodynamic interactions between two spherical particles},
volume = {10},
journal = {PhysicoChem. Hyd.}
}

@article{Masoud.Stone2019,
  author = {Masoud, Hassan and Stone, Howard A.},
  title = {The reciprocal theorem in fluid dynamics and transport phenomena},
  journal = {J. Fluid Mech.},
  year = 2019,
  volume = {879},
  pages = {P1},
  doi = {10.1017/jfm.2019.553},
}

@article{Lorentz,
  title={Eene algemeene stelling omtrent de beweging eener vloeistof met wrijving en eenige daaruit afgeleide gevolgen},
  author= {Lorentz, H. A.},
  journal = {Zittingsverslag van de Koninklijke Akademie van Wetenschappen te Amsterdam},
  volume = {5},
  pages = {168},
  year={1896},
  doi = {10.1007/978-94-009-0225-1_2}
}

@book{Lamb:1924,
  title={Hydrodynamics},
  author={Lamb, Horace},
  year={1924},
  publisher={Cambridge University Press, Cambridge}
}



\end{document}